\documentclass[runningheads,a4paper]{llncs}

\usepackage{amssymb}
\usepackage{graphicx}
\usepackage{mathtools}
\usepackage{stmaryrd}
\def\multiset#1{\ensuremath{\left[\kern-.3em\left[#1\right]\kern-.3em\right]}}
\usepackage{hyperref}
\usepackage{breakurl}
\urldef{\mailsa}\path|alexei@behold-research.com|

\begin{document}

\mainmatter  

\title{Memantic: A Medical Knowledge Discovery Engine}

%
%
\author{Alexei Yavlinsky}
\authorrunning{Memantic: A Medical Knowledge Discovery Engine}

\institute{Behold Research Limited,\\
Suite 266, 2 London Bridge Walk, London, SE1 2SX, UK\\
\mailsa\\
}
%
%

\toctitle{Memantic: A Medical Knowledge Discovery Engine}
		\tocauthor{Alexei Yavlinsky}
\maketitle

\begin{abstract}
We present a system that constructs and maintains an up-to-date 
co-occurrence network of medical concepts based on continuously 
mining the latest biomedical literature. Users can explore this 
network visually via a concise online interface to quickly 
discover important and novel relationships between medical entities. 
This enables users to rapidly gain contextual understanding of their 
medical topics of interest, and we believe this 
constitutes a significant user experience improvement over contemporary 
search engines operating in the biomedical literature domain.
\end{abstract}

\section{Introduction}

Progress in the development of search engines and digital libraries in the 
last two decades has revolutionised access to biomedical literature. 
One only needs to recall that before the advent of PubMed and Google
researchers had to mainly use library index cards to identify articles of interest.

Besides giving instant access to papers based on keyword queries, search 
engine systems enable users to browse citation information, thus offering
a much more efficient process for performing background research and literature reviews.
Google Scholar also provides authorship statistics that allow for quick 
identification of influential and prolific researchers.

Nonetheless, keyword-based searches and citation analysis have their limitations. Examples
	of medical information needs that cannot be fulfilled directly by the current generation of search engines include
	``\emph{give me the list of all diseases and syndromes that are comorbid or are otherwise related to asthma}''
or ``\emph{give me all pharmacological agents that have been used in the treatment of migraine}''. Whereas the latter type of query is
									self-explanatory, the former is significant because asthma is a complex and multifactorial disease. Thus, the co-occurring conditions may share common aetiologies with one or more of its factors or may otherwise shed light on its causes, providing researchers and practitioners with additional avenues for understanding the disease and exploring new treatment options. While there are online resources that can give partial answers to such questions (examples include WebMD, Medscape, The Mayo Clinic and Wikipedia), these are curated manually and therefore may not always be up-to-date. The alternative is to supply broad queries to a search engine of choice, meticulously browsing and analysing the returned results to build the desired big picture. The results will be organised according to the search engine's global ranking criteria, which, from the user's perspective, are arbitrary with respect to the actual information need. This makes the browsing process unnecessarily time-consuming.

The above limitations have been apparent to the community for quite some\\ time. In his 2002 paper ``Mining the bibliome: searching for a needle in a haystack" \cite{grivell:2002}, Les Grivell outlined a number of then-unmet challenges for biomedical search engines, amongst them: \begin{itemize}
	
	\item \emph{Reducing the number of search results to a manageable level:} users normally address this by trying many different Boolean search term combinations to find the query refinement that frames their original query in the desired context.
	
	\item \emph{Handling synonymy and polysemy of biomedical search terms:} to lower the risk of missing relevant search results, users must apply their erudition and issue separate queries for each of the search terms' synonyms. On the other hand, term polysemy further complicates the first problem mentioned above.

	\item \emph{Extracting meaningful information from articles:} the freedom of natural language makes it difficult to extract relevant biomedical terms unambiguously, thus reducing search accuracy.
\end{itemize} To address these shortcomings, the author suggested visually structuring search results using contextual information facilitated by controlled vocabularies and biomedical ontologies. However, in the twelve years after Grivell's paper was published there has been little indication that popular search engines are heeding his advice, while the amount of medical literature has since more than doubled \--- from 11 million at the time of his writing to almost 25 million articles today\footnote{As indexed by PubMed at the time of publication} \--- and continues to grow at an exponential rate, making Grivell's suggestions ever more pertinent.

\begin{figure}
\centering
\includegraphics[width=0.6\textwidth]{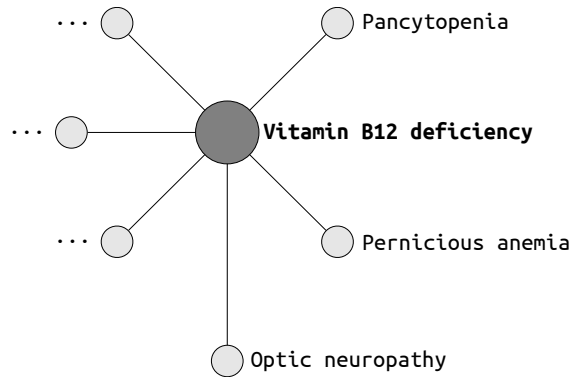}
\caption{Memantic's visualisation of diseases related to \emph{vitamin B12 deficiency}.}
\label{fig:figure1}
\end{figure}

In this paper we present a novel search engine for medical publications \--- called ``Memantic" \--- that aims to tackle the challenges described above. Memantic captures relationships between medical concepts by mining biomedical literature and organises these relationships visually according to a well-known medical ontology \cite{lowe:1994}. To give an example, a search for ``Vitamin B12 deficiency" will yield a visual representation of all related diseases, symptoms and other medical entities that Memantic has discovered from the 25 million medical publications and abstracts mentioned above, as well as a number of medical encyclopaedias. Figure~\ref{fig:figure1} shows a conceptual visualisation of this idea.

\begin{figure}
\centering
\includegraphics[width=1.0\textwidth]{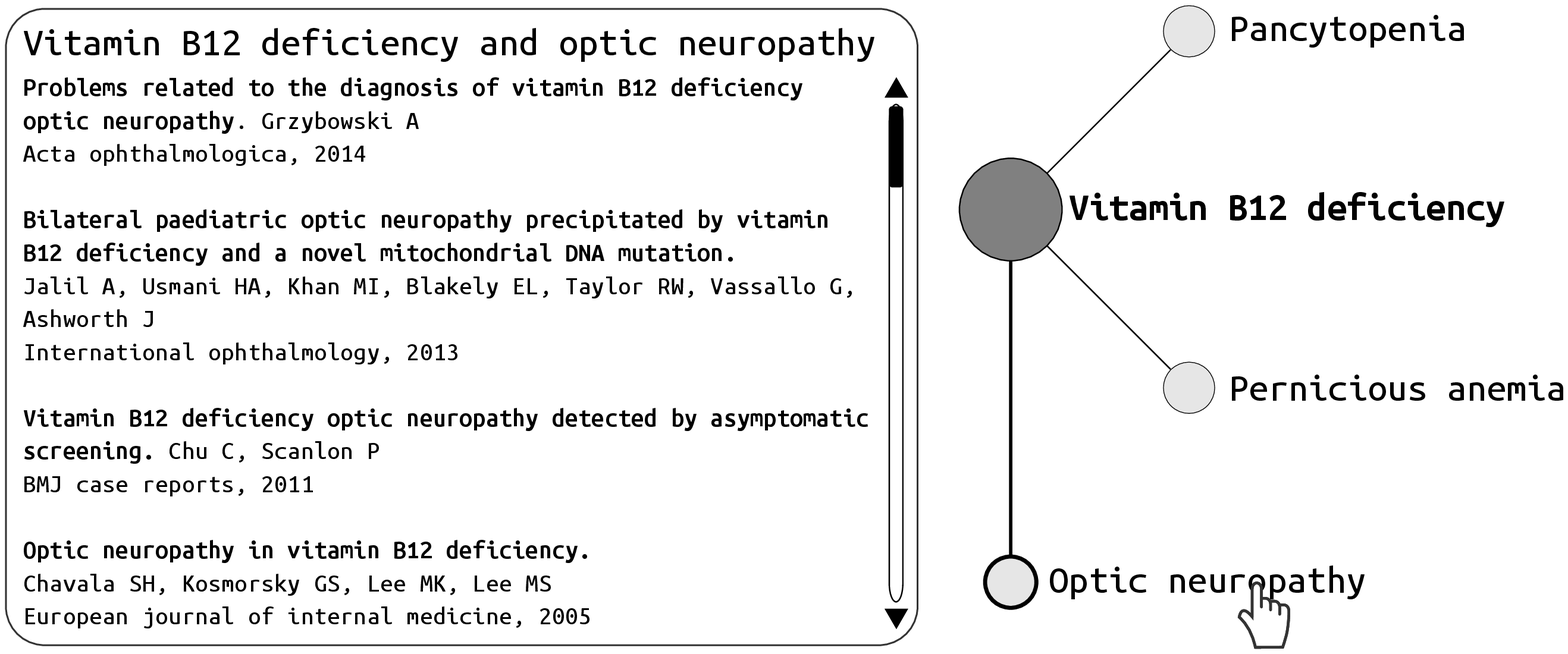}
\caption{Exploring the relationship between \emph{vitamin B12 deficiency} and \emph{optic neuropathy}.}
\label{fig:figure1_5}
\end{figure}

\begin{figure}
\centering
\includegraphics[width=0.7\textwidth]{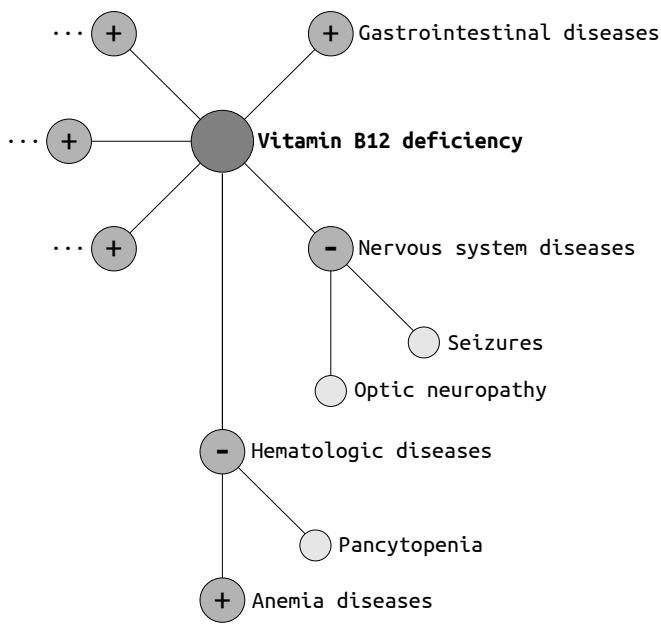}
\caption{Hierarchically grouping diseases related to \emph{vitamin B12 deficiency}.}
\label{fig:figure2}
\end{figure}

The user can explore a relationship of interest (such as the one between ``Vitamin B12 deficiency'' and ``optic neuropathy", for instance) by clicking on it, which will bring up links to all the scientific texts that have been discovered to support that relationship (see Figure~\ref{fig:figure1_5}). 
Furthermore, the user can select the desired type of related concepts  \--- such as ``diseases", ``symptoms", ``pharmacological agents", ``physiological functions", and so on \--- and use it as a filter to make the visualisation even more concise. Finally, the related concepts can be semantically grouped into an expandable tree hierarchy to further reduce screen clutter and to let the user quickly navigate to the relevant area of interest (Figure~\ref{fig:figure2}). 

We believe Memantic can save a considerable amount of time for researchers, medical students and practitioners who are investigating a particular condition, symptom or drug by giving a quick yet rich ``mind map"-like overview of related medical concepts. We call our system a \emph{knowledge discovery engine} because in contrast to traditional search systems it allows the user to quickly identify relationships that were previously unfamiliar to them. We achieve this in two ways:
\begin{itemize}
	
	\item \emph{Concisely organising related medical entities without duplication:} Memantic first presents all medical terms related to the query concept and then groups publications by the presence of each such term in addition to the query itself. The hierarchical nature of this grouping allows the user to quickly establish previously unencountered relationships and to drill down into the hierarchy to only look at the papers concerning such relationships. Contrast this with the same search performed on Google, where the user normally gets a number of links, many of which have the same title; the user has to go through each link to see if it contains any novel information that is relevant to their query.
	
	\item \emph{Keeping the index of relationships up-to-date:} Memantic perpetually renews its index by continuously mining the biomedical literature, extracting new relationships and adding supporting publications to the ones already discovered. The key advantage of Memantic's user interface is that novel relationships become apparent to the user much quicker than on standard search engines. For example, Google may index a new research paper that exposes a previously unexplored connection between a particular drug and the disease that is being searched for by the user. However, Google may not assign that paper the sufficient weight for it to appear in the first few pages of the search results, thus making it invisible to the people searching for the disease who do not persevere in clicking past those initial pages.  
	
\end{itemize} The above points are illustrated in detail in Section~\ref{sec:cases} with a number of real life examples.

The inspiration for our approach comes from the work on word co-occurrence networks done in the early 1980s \cite{callon:1983}. There, the authors set out to create a map of relationships between  scientific concepts across a number of disciplines. To do this, they compile a vocabulary of key scientific terms, turn them into graph vertices, and create edges between any two such vertices for which the corresponding terms are found to co-occur in scholarly texts under a set of threshold criteria (such as the frequency of their co-occurrence). Finally, they construct a graph of relationships between such terms by applying this procedure to a large database of relevant literature. This graph is then visualised in two dimensions and can be studied by researchers to explore relationships between key scientific concepts, both within and across a number of different scientific domains. Our interpretation of this idea benefits from the use of a thoroughly curated medical ontology called Medical Subject Headings (MeSH) \cite{lowe:1994}, where the relevant keywords for biomedical concepts are well established and are hierarchically categorised, and where synonymous terms are grouped together using unique identifiers. 

The rest of this paper is structured as follows. Section 2 gives an overview of related work. Section 3 describes the algorithm for constructing the co-occurrence network of medical concepts and our user interface design for exploring it. We demonstrate the usefulness of our search engine in Section 4 with a number of case studies and conclude this paper with an outline of potential future work in Section 5.


\section{Related work} 

\subsection{Network medicine}

Network medicine links experimental data on gene, protein, metabolic interactions with clinical knowledge about diseases into interaction networks. These networks can be studied to  enhance the understanding of diseases and their treatments. Such networks offer a bird's eye view of the multiple factors that can impact a particular disease and lend themselves to computational and mathematical analysis that can help identify novel disease pathways and predict patient drug response. A good review of these methods can be found in \cite{jacunski:2013}. An illustrative example of this idea is the work by Goh et al. \cite{goh:2007}, where the authors construct a network of human disorders and disease genes linked by known disorder\--gene associations, which they term a ``diseasome''. They then visualise this network as a graph, and use it to identify sets of genes responsible for multiple medical conditions. Interestingly, they choose to manually categorise each disorder into 20 primary disorder classes \--- based on the physiological system affected by each disorder \--- instead of using labels provided by a medical ontology such as MeSH.

Memantic differs from the above approach because it does not attempt to link medical concepts via shared gene or protein interactions. Instead, two concepts are linked only when the potential relationship between them has already been explicitly highlighted in scientific publications. We believe this makes the resulting network more useful in the clinical setting where medical specialists need existing research to support diagnoses and the associated treatments.

\subsection{Clinical decision support systems}

A clinical decision support system (CDSS) is a software system that is designed to assist healthcare professionals in clinical decision making. Research in this field goes back to the early 1970s \cite{weiss:1974} and one of the earliest successful implementations of this concept is the CASNET/Glaucoma system that uses the manually supplied expertise of glaucoma specialists to construct a causal model of the disease \cite{weiss:1978}. This model relates disease states (e.g. \emph{``open angle glaucoma"} vs. \emph{``acute closure glaucoma"}) with clinical observations (e.g. \emph{``reduced visual acuity"} or \emph{``dilated pupil"}) via a causal network of corresponding pathophysiological states (\emph{``angle closure"} leading to \emph{``elevated intraocular pressure"}). The therapist interacts with the system through a ``consultation program'' by answering a series of questions about the patient's medical background and presentation of the disease. The system's inference engine then uses the disease model to map the supplied answers to the potential causes of the patient's condition and the recommended therapies. CASNET/Glaucoma has been able to achieve a high level of competence in analysing complex cases of the disease.

Other notable systems originating from research done in the 1970s include MYCIN \cite{shortliffe:1975,shortliffe:1976}, designed to diagnose and recommend treatments for a number of blood infections (such as bacteremia or meningitis), ONCOCIN \cite{shortliffe:1981}, built to assist physicians with the treatment of cancer patients receiving chemotherapy and INTERNIST I \cite{pople:1975,miller:1982}, used for the diagnosis of complex problems in general internal medicine. DXplain \cite{barnett:1987}, another clinical decision support system developed in the early 1980s at the Laboratory of Computer Science, Massachusetts General Hospital, has since become an online service with both educational and clinical uses \cite{dxplain:online}. The system utilises a set of clinical findings to produce a ranked list of diagnoses which might explain or otherwise be associated with the clinical manifestations. DXplain contains 2,200 diseases and 5,000 symptoms in its knowledge base.

\subsubsection{Map of Medicine.}

Map of Medicine is a CDSS that is currently used by the UK National Health Service \cite{mapofmed:online}. Information pertaining to a particular health condition is visualised as a `care map', which is a flow chart made up of a series of steps to be taken with the patient, such as tests, treatments, referral to specialists, and links to other care maps. These maps are manually curated and kept up-to-date by a pool of medical specialists, who use the latest available medical evidence for this purpose. A healthcare practitioner can use these maps as a guide on how to tackle patients' health problems.

\subsubsection{WatsonPaths.}

Like the Map of Medicine, IBM's WatsonPaths visualises medical information using flow charts. However, instead of building on the manual efforts of medical curators, WatsonPaths constructs such flow charts \--- or ``paths'' \--- automatically, by scanning medical literature. These can be used for managing a previously identified medical condition and for establishing a set of possible diagnoses based on observed symptoms and laboratory test results. To quote IBM's website: ``when presented with a medical case, [WatsonPaths] extracts statements based on the knowledge it has learned from being trained by medical doctors and from medical literature. Using Watson's question-answering abilities, WatsonPaths can examine the scenario from many angles, working its way through chains of evidence \--- pulling from reference materials, clinical guidelines and medical journals in real-time \--- and drawing inferences to support or refute a set of hypotheses. This ability to map medical evidence allows medical professionals to consider new factors that may help them to create additional differential diagnosis and treatment options. [...] WatsonPaths incorporates feedback from the physician who can drill down into the medical text to decide if certain chains of evidence are more important, provide additional insights and information, and weigh which paths of inferences the physician determines lead to the strongest conclusions.'' \cite{watpaths:online}. As WatsonPaths was not publicly available at the time of writing, we could not review 	these features in greater depth. However, it is reported that the system is currently undergoing trials in a number of medical schools \cite{watpaths:schools}.

\subsubsection{Google Health Cards.}

As of 2013, when a user issues a medical query, such as \emph{``causes of multiple sclerosis''} or \emph{``symptoms of the flu"}, Google shows short extracts from authoritative sites like WebMD and MayoClinic that give summarised answers to the medical question that is presumed to be contained within the query \cite{ghealthcards:online}. However, Google expressly disclaims responsibility for their accuracy.

\subsubsection{Memantic in the context of clinical decision support systems.}

Unlike clinical decision support systems, Memantic does not suggest diagnoses based on symptoms and other observations. Instead, it simply exposes relationships between medical concepts, leaving it up to the healthcare practitioner to choose how to use any novel information they discover for the medical case at hand. One important aspect of our system is that we do not automatically infer causality for any such relationship, instead encouraging the user to read the supporting literature to establish any possible causal aspects for themselves. Our philosophy is that the doctor should be in charge at all times and should not be unduly influenced by automated suggestions. Instead of prescribing diagnoses or specific courses of action, Memantic offers a quick way to broaden the practitioner's horizons for a particular medical topic by highlighting relevant relationships that might otherwise be easily overlooked. 

\subsection{Document and term clustering systems}

Grouping related terms and documents for improving search accuracy is a well-established approach in information retrieval. \emph{Latent Semantic Indexing} (LSI) was the first formal framework for this purpose \cite{deerwester:1990}. Within this framework, a linear-algebraic technique called singular value decomposition is applied to the matrix that represents the occurrence of terms within a collection documents, and the results are used to identify groups of terms that frequently co-occur. The underlying assumption is that words that often appear together are likely to have similar meanings (or, to put it in a different way, are ``\emph{semantically similar}''). LSI offers a principled way of querying a document space such that it becomes possible to retrieve the documents containing terms that are semantically similar to those in the query. Therefore, if one searches for the term \emph{`car'}, the documents that have the word \emph{`automobile'} will be ranked highly in the returned list of search results, even if those documents do not contain the term \emph{`car'} itself. However, such relationships between words are captured implicitly by singular value decomposition and are not made known to the user. Thus, the latter is left unaware of why a particular set of documents is ranked highly in response to his or her query. Conversely, Memantic explicitly exposes the captured relationships between medical concepts by letting the user interact directly with the co-occurrence network.

Vivisimo \cite{koshman:2006}, a document clustering system developed in the early 2000s, partially addresses the implicit nature of LSI by hierarchically clustering search results based on the semantic similarity of the returned documents. Each cluster is assigned a textual label that corresponds to the most representative term out of those used for grouping the documents within that cluster. The cluster hierarchy is visualised as a one-dimensional, expandable tree, where each intermediate tree node represents a document cluster. The tree leaves are the actual documents belonging to one or more clusters. This approach is much closer to Memantic in its spirit than is LSI. However, since Vivisimo has to handle arbitrary queries that are not domain specific, it has to rely on automatic techniques for finding representative cluster labels, which can give rise to errors. In the case of Memantic, such errors are prevented by the use of the MeSH ontology, which ensures that only medically relevant terms are used for grouping documents together.

\section{System description}

\subsection{Co-occurrence network construction}

	Consider a dictionary set of medical terms $\{m \in M\}$ and a database of medical documents $\{d \in D\}$. Denote $d^t$ as the title of document $d$, $d^a$ as its abstract and $d^f$ as the document's full text body, each represented as a string of text. For each available document $d_i$, identify in $d_{i}^{t}$, $d_{i}^{a}$ and $d_{i}^{f}$ the longest non-overlapping substrings that match medical terms in $M$. Denote those as multisets  $\llbracket n \text{ in } T_{d_i} \rrbracket$, $\llbracket r \text{ in } A_{d_i} \rrbracket$ and $\llbracket b \text{ in } F_{d_i} \rrbracket$, respectively. Further denote $T\text{\#}t$ as the number of times the term $t$ occurs in the multiset $T$ and $T^{\prime}$ as the set of all terms in $T$. Defines the co-occurrence operator between two multisets of extracted medical terms as
\[	
	P \circ Q = \left\{ (p,q,z_{P,Q}(P{\#}p,Q{\#}q)) : p \in P^{\prime}, q \in Q^{\prime} \right\}
\]
The result of this operator is a set of tuples corresponding to all possible ordered pairs of terms, whose first component is a term that occurs at least once in $P$, the second is a term that similarly occurs in $Q$ and the third is the real-valued output of the weight function $z_{PQ}$ that takes into account the number of times the first term appears in $P$ and the second in $Q$. Compute the co-occurrence sets for $T_{d_i} \circ T_{d_i}$, $T_{d_i} \circ A_{d_i}$, $T_{d_i} \circ F_{d_i}$, $A_{d_i} \circ A_{d_i}$, $A_{d_i} \circ F_{d_i}$ and $F_{d_i} \circ F_{d_i}$. Define a mapping function 
\[
f : M \mapsto \{ x \mid x \le \left\vert{M}\right\vert, x \in \mathbb{N}\}
	\] such that every term in the medical dictionary maps to a unique integer $x$ that can range between 1 and the size of the dictionary. Further define the co-occurrence matrix as the real-valued, $\left\vert{M}\right\vert \times \left\vert{M}\right\vert$ matrix $\mathbf{C}$. Given a co-occurrence set $P \circ Q$, for every tuple $(p_j, q_j, z_j)$ in the set, increment $\mathbf{C}(f(p_j),f(q_j))$ by $w_{P,Q}z_j$, where $w_{P,Q}$ is the weight coefficient for $P \circ Q$. Apply this step for every co-occurrence set computed above. Repeat the entire procedure for every document $d \in D$, such that $\mathbf{C}$ becomes the matrix of relationships between the medical terms in $M$, where $\mathbf{C}(f(m_k),f(m_l))$ represents the `relatedness' score between the terms $m_l$ and $m_k$.

In our original implementation of Memantic 
\[
\forall{p \in P^{\prime}, q \in Q^{\prime}}, {z_{P,Q}(P{\#}p,Q{\#}q) = 1}
\] and the weight coefficients $w$ were determined heuristically such that each co-occurrence set type (e.g. `title\--abstract') would have a corresponding constant weight value across all documents.

In simple terms, $\mathbf{C}$ defines our co-occurrence network and indicates how often pairs of medical terms occur together in medical documents. The weighting function $z$ and the weighting coefficients $w$ are used to rank the relative importance of two terms occurring together in the same title of a document versus e.g. one occurring in the document's title and the other in the abstract of the same document, before computing the final relatedness score.

\subsection{Document database and medical dictionary}

To construct the above co-occurrence network, we used the entire catalogue of medical publications available via PubMed \cite{pubmed:online} (approximately 25 million at the time of writing). We analysed the articles' full text where the former were accessible via PubMed's OpenAccess policy and only their metadata otherwise (title and abstract). Additionally, we indexed web pages about medical conditions from the Medscape online medical encyclopaedia \cite{medscape:online}. We used the National Institute of Health's MeSH ontology as the source of entries in our medical dictionary, utilising both the main descriptors and the supplementary concept records available therein. We utilised MeSH unique identifiers as our medical terms to ensure that synonymous entries in the dictionary map to the same underlying medical concepts in the co-occurrence network.

\subsection{Visualisation and user interface}

\begin{figure}
\centering
\includegraphics[width=1\textwidth]{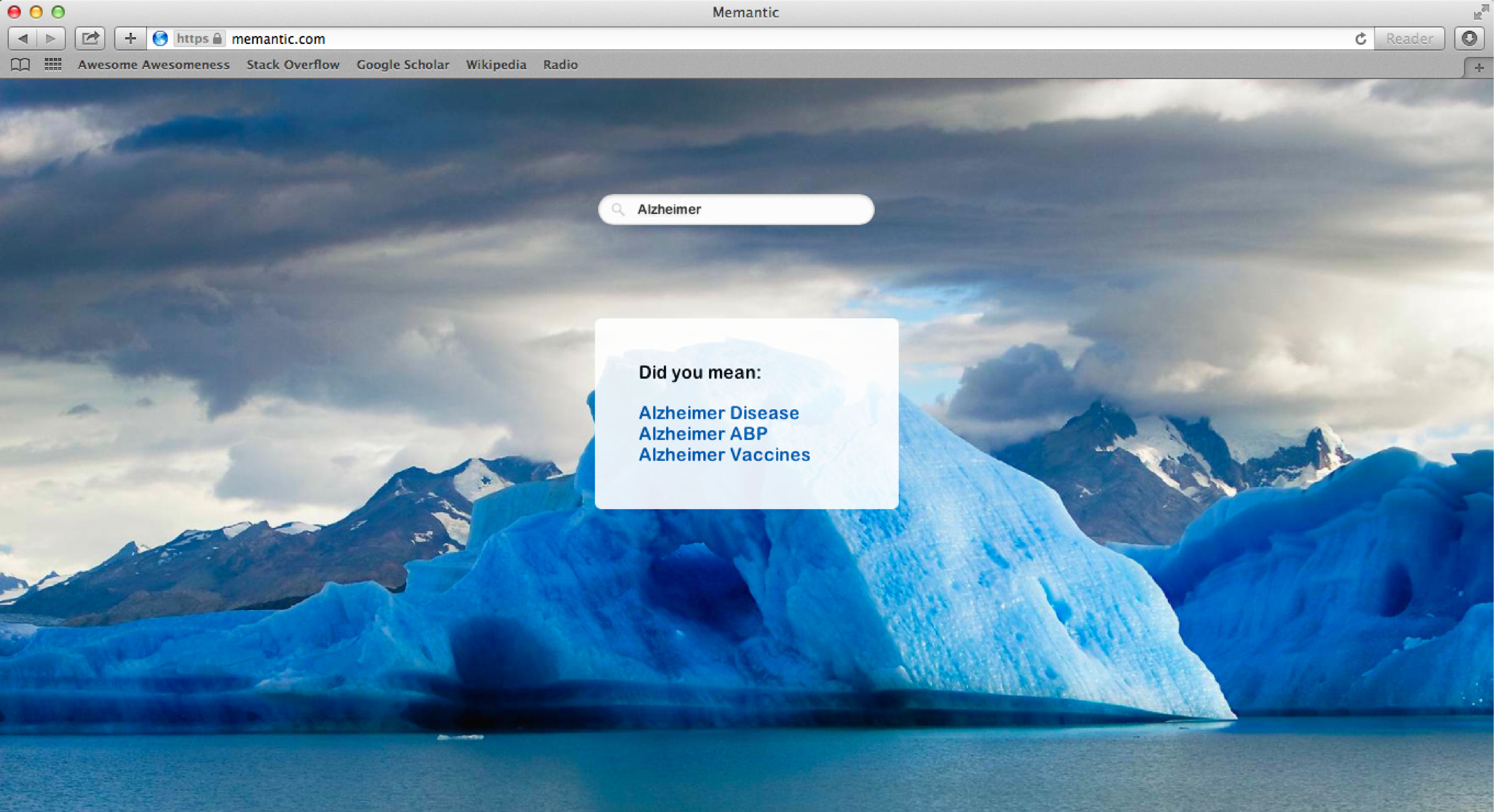}
	\caption{Query interface, with a list of spelling-based suggestions for the ambiguous query `\emph{Alzheimer}'.}
\label{fig:query}
\end{figure}

\subsubsection{Querying.} The user starts with a query containing a single medical concept. This can be any term that is contained in the MeSH controlled vocabulary, and is not restricted to diseases or symptoms. For example, a drug name or a geographical region can also be supplied. If the query is not present in MeSH, Memantic will offer a list of suggestions that are similarly spelled, ranked by the Levenshtein distance between the query term and the suggested vocabulary entries (Figure \ref{fig:query}).

\begin{figure}
\centering
\includegraphics[width=1\textwidth]{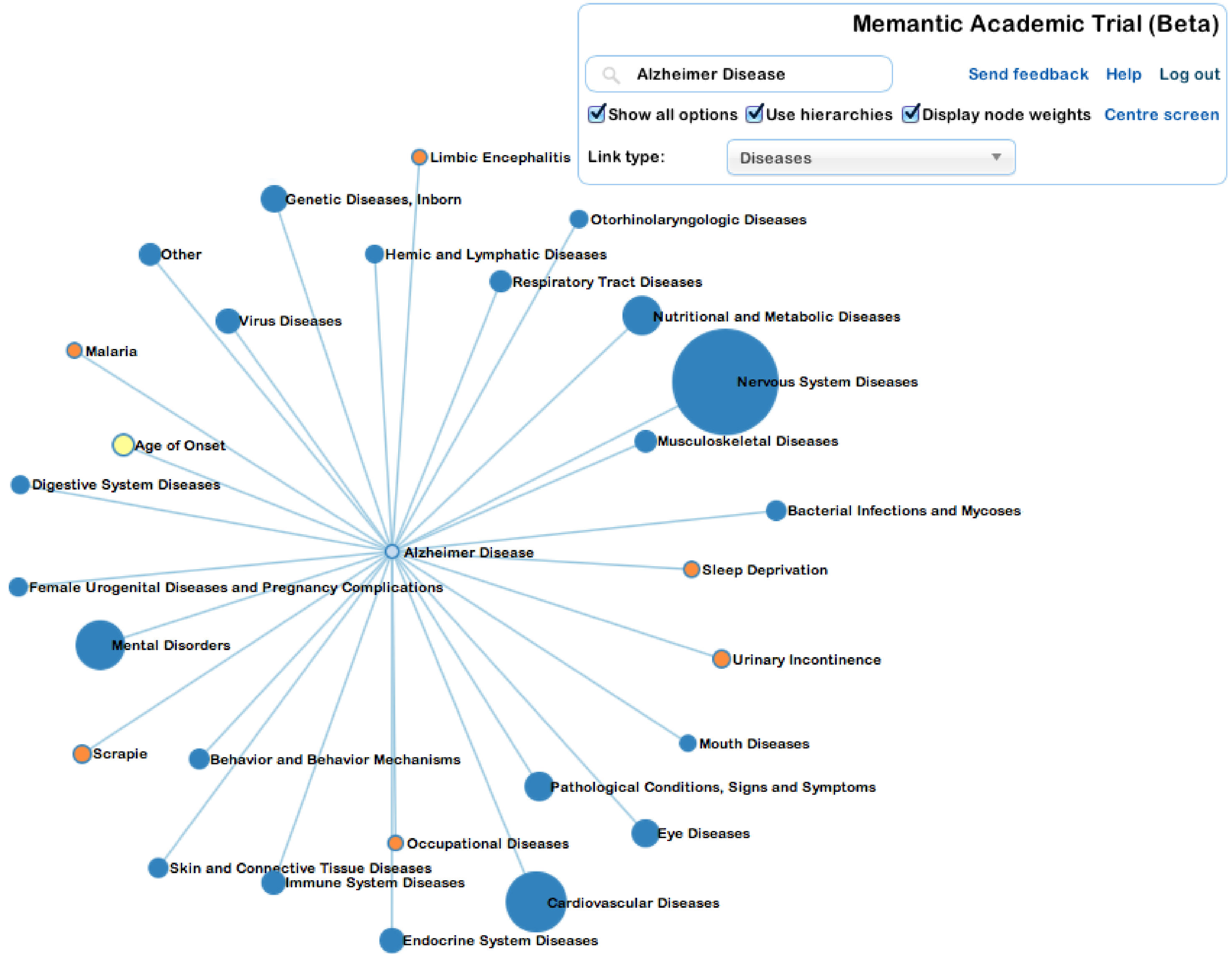}
\caption{Hierarchically organised search results for the query in Figure \ref{fig:query}.}
\label{fig:result_alz}
\end{figure}

\subsubsection{Search results.} 
\begin{figure}
\centering
\includegraphics[width=1\textwidth]{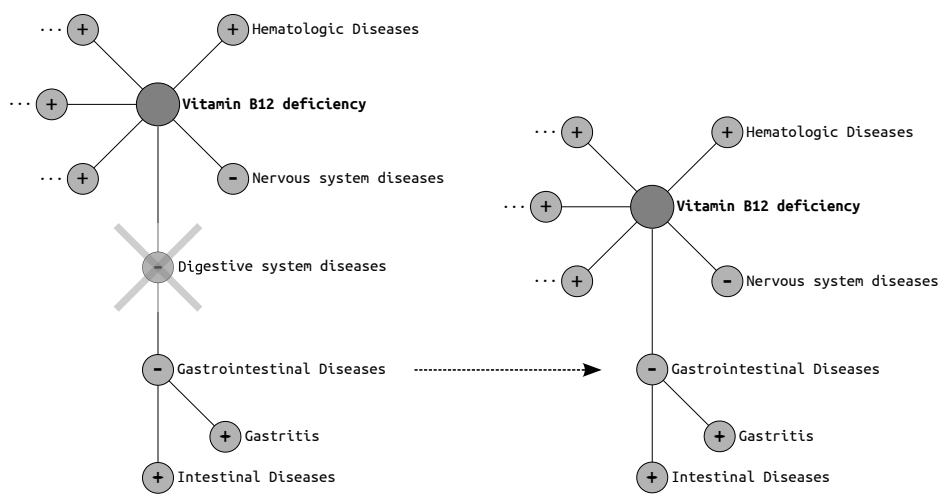}
	\caption{Removing an intermediate node with a single child from the tree structure prior to visualisation.}
\label{fig:figure3}
\end{figure}

\begin{figure}
\centering
\includegraphics[width=0.85\textwidth]{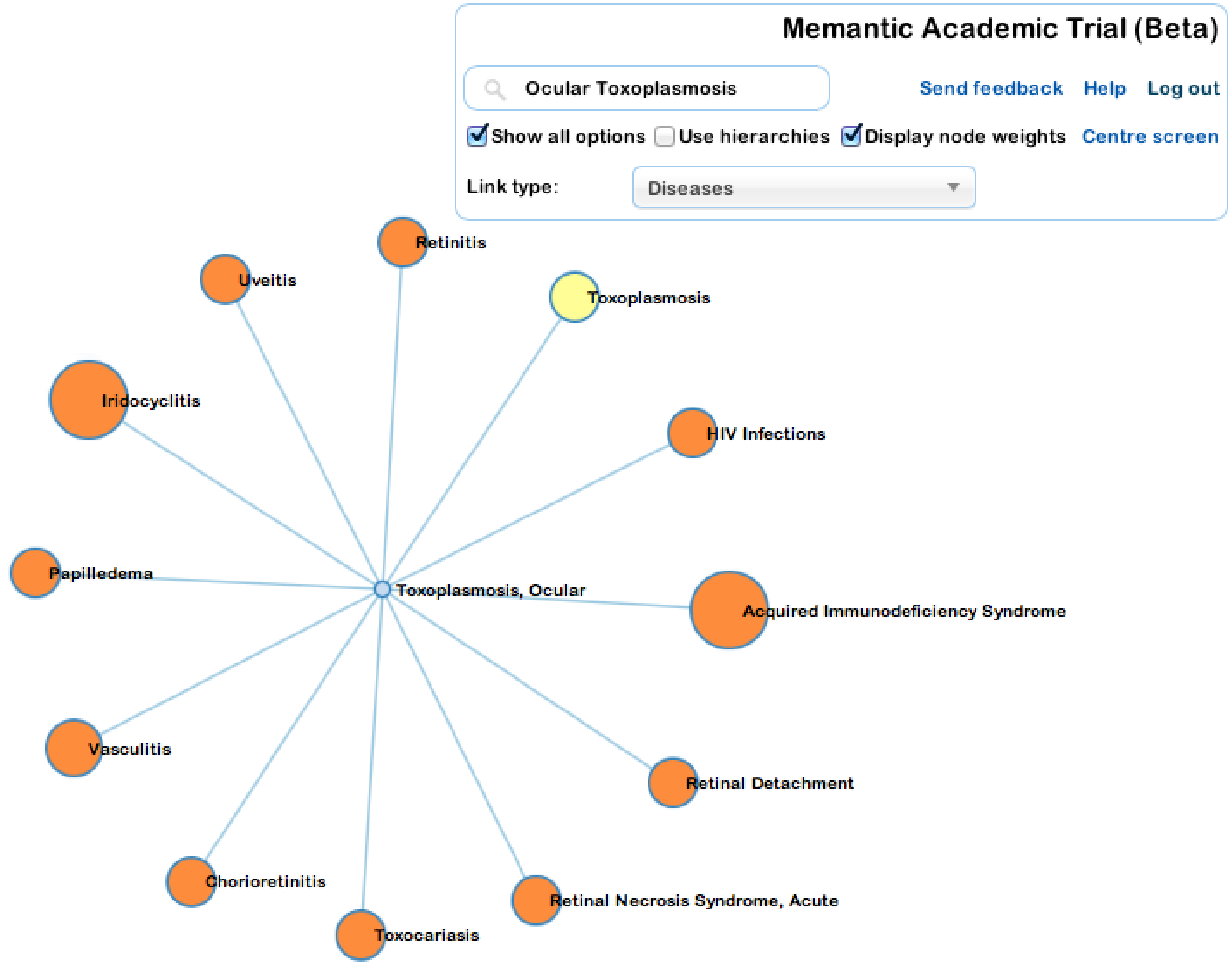}
\caption{Flat result visualisation for the query \emph{``Ocular toxoplasmosis''}. This is a convenient way to display concept connections when there aren't too many of them.}
\label{fig:result_toxo}
\end{figure}

\begin{figure}
\centering
\includegraphics[width=1\textwidth]{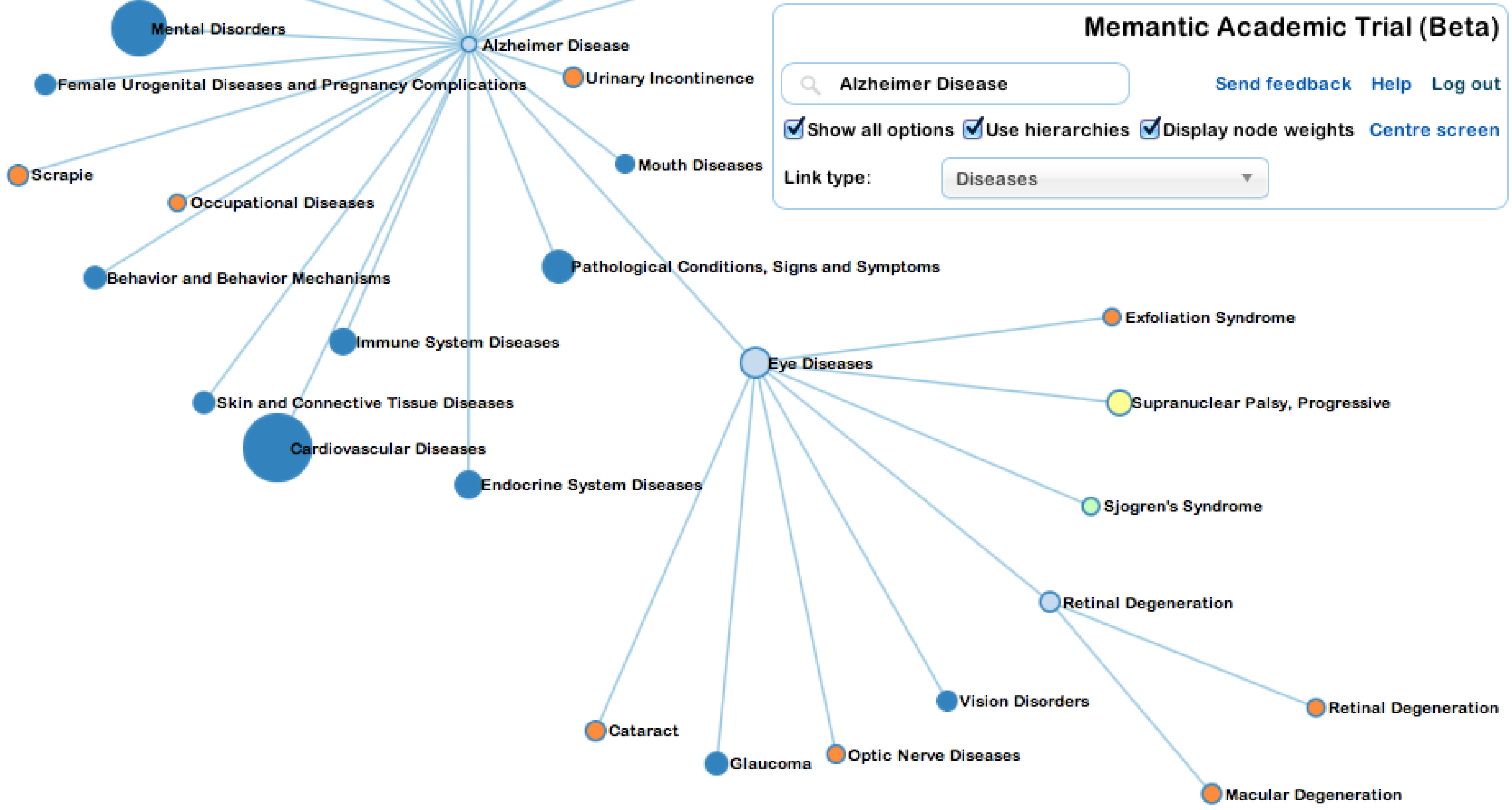}
\caption{User expanding the tree presented in Figure \ref{fig:result_alz}. Expanding a collapsed intermediate node changes its colour from dark to light blue.}
\label{fig:result_alz_expand}
\end{figure}

Search results are visualised as a two-dimensional tree, which is centred on the tree root. The latter represents the query concept and the tree leaves are the concepts connected to the query in the co-occurrence network\footnote{Memantic displays a concept only when it occurs together with the query term in at least two different research articles or in at least one medical encyclopaedia entry that directly concerns either that concept or the query term itself. This is done to filter out spurious co-occurrences that are not scientifically significant.}. The tree is rendered using the d3.js Collapsible Force Layout \cite{d3:online}, which employs an optimisation algorithm to evenly position the tree nodes on the screen. The nodes can be dragged around and repositioned by the user to maximise the legibility of the node labels. There are two modes of visualising the tree: \emph{hierarchical} and \emph{flat}:

\begin{itemize}
	
\item In the first mode, intermediate tree nodes (coloured blue) are used to hierarchically group leaf nodes according to their conceptual categories (Figure \ref{fig:result_alz}). The set of hierarchical categories of a leaf node is based on the MeSH hierarchy descriptor of the node's concept. An intermediate node can be collapsed by clicking on it, which will hide all the descending nodes from view. When the search results are returned, all intermediate nodes of depth one and greater are collapsed by default. This is done to reduce screen clutter and can be especially useful when the query concept has many connections. The user can then expand the nodes that correspond to the categories of interest, and do so progressively until the desired set of connections is revealed (Figure \ref{fig:result_alz_expand}). If an intermediate node
has one child node,
it is automatically excised from the tree in the manner shown in Figure \ref{fig:figure3}. This procedure is applied repeatedly until there are no such nodes left in the tree. This is done to reduce the effort of navigating through the tree structure.

\item In the second mode, all the leaves are directly connected to the root node (Figure \ref{fig:result_toxo}). It is possible to toggle between the two modes by clicking the ``use hierarchies" checkbox in the floating toolbox located in the top right hand corner of the screen.

\end{itemize}

\subsubsection{Leaf nodes.} 
\begin{figure}
\centering
\includegraphics[width=1\textwidth]{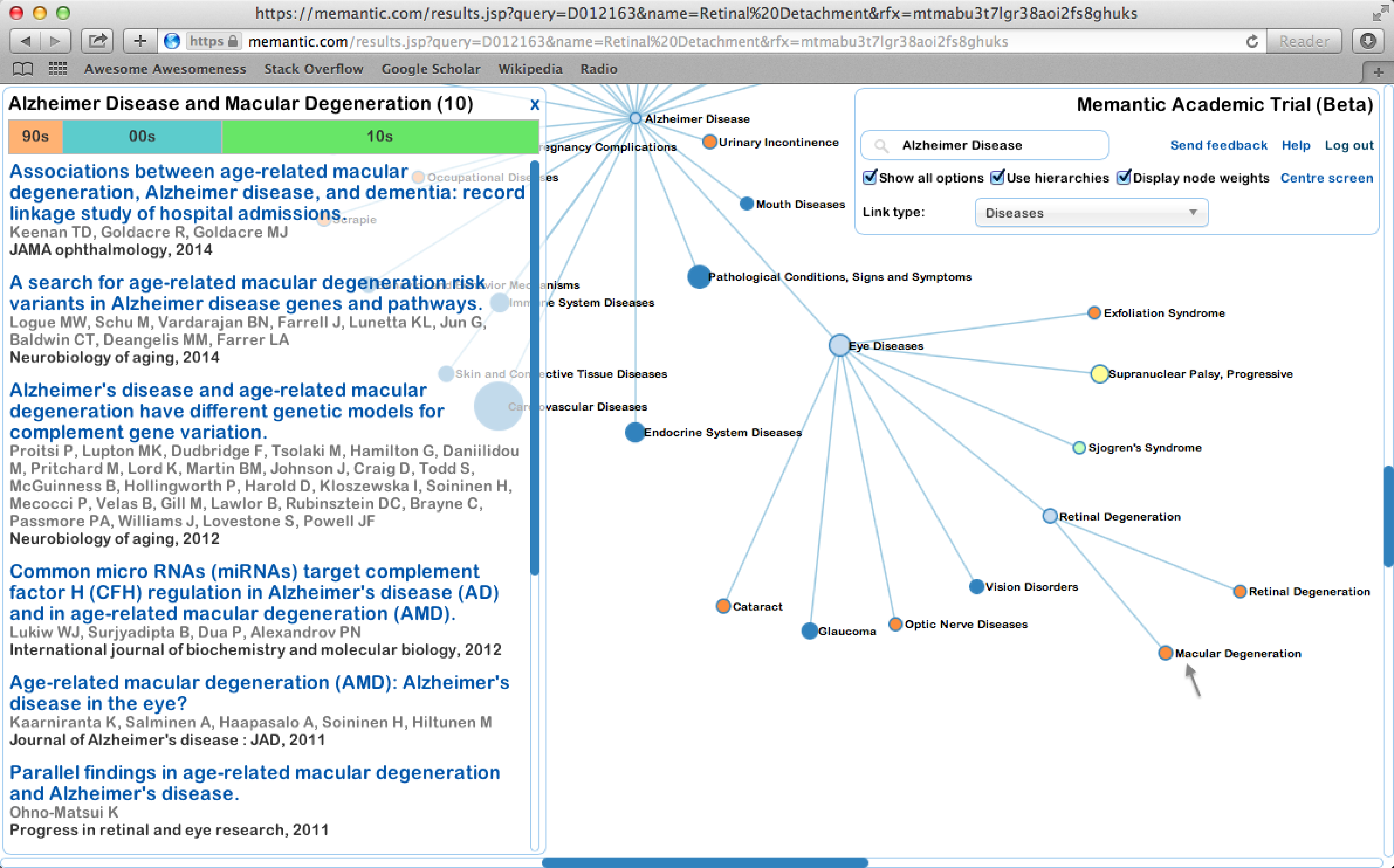}
\caption{User clicking on a concept associated with the query in Figure \ref{fig:result_alz}. In this case, the user is exploring the connection between \emph{Alzheimer disease} and \emph{macular degeneration}. The number in brackets is the total count of all relevant articles returned in the publication list.}
\label{fig:result_alz_pubs}
\end{figure}

Clicking on the text of a leaf node will unveil a panel on the left side of the screen, listing links to scientific articles that contain the association between the leaf's concept and that of the root node (Figure \ref{fig:result_alz_pubs}). The articles are ordered by publication date, starting with the most recently published paper. The top of the publication panel contains a horizontally stacked bar chart that gives a visual breakdown of the number of relevant publications by each decade. Clicking on a decade bar will scroll the publication list to the last publication from that decade.

\begin{figure}
\centering
\includegraphics[width=1\textwidth]{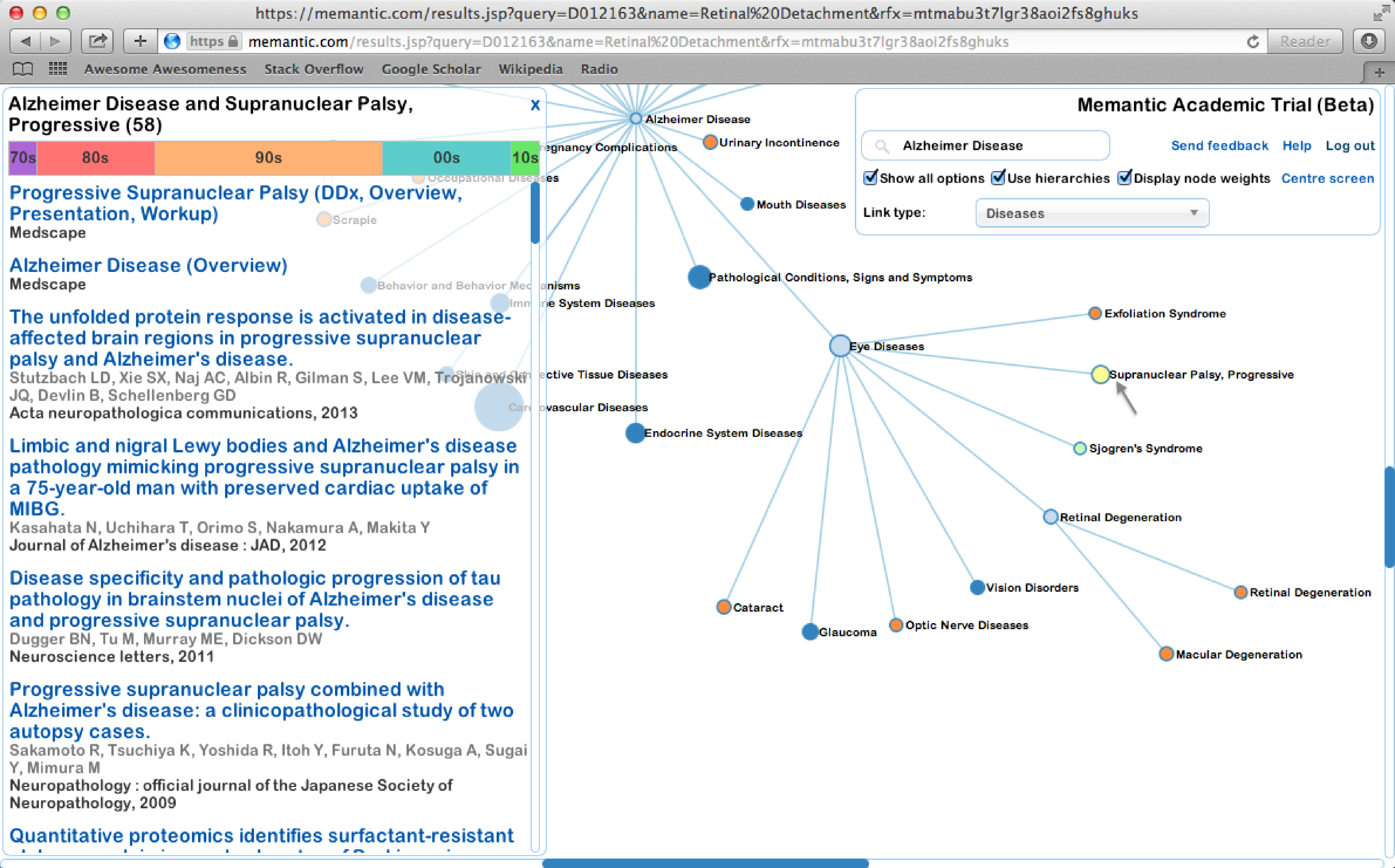}
\caption{User clicking on a concept associated with the query in Figure \ref{fig:result_alz}. Now the user is exploring the association between \emph{Alzheimer disease} and \emph{progressive supranuclear palsy}. Note the medical encyclopaedia links that precede the list of relevant scientific publications.}
\label{fig:result_alz_pubs_enc}
\end{figure}

If the association has also been found in one of the medical encyclopaedias indexed by Memantic, a link to the corresponding encyclopaedia article will be displayed in the publication panel above the list of scientific articles (Figure \ref{fig:result_alz_pubs_enc}).

Clicking on the circle part of the leaf node will issue a new search query with the text of the node's label, which, in effect,``re-centres" the user interface on the clicked concept. This allows the user to explore the co-occurrence network by jumping from one concept to the next.

\subsubsection{Link type.} 
\begin{figure}
\centering
\includegraphics[width=1\textwidth]{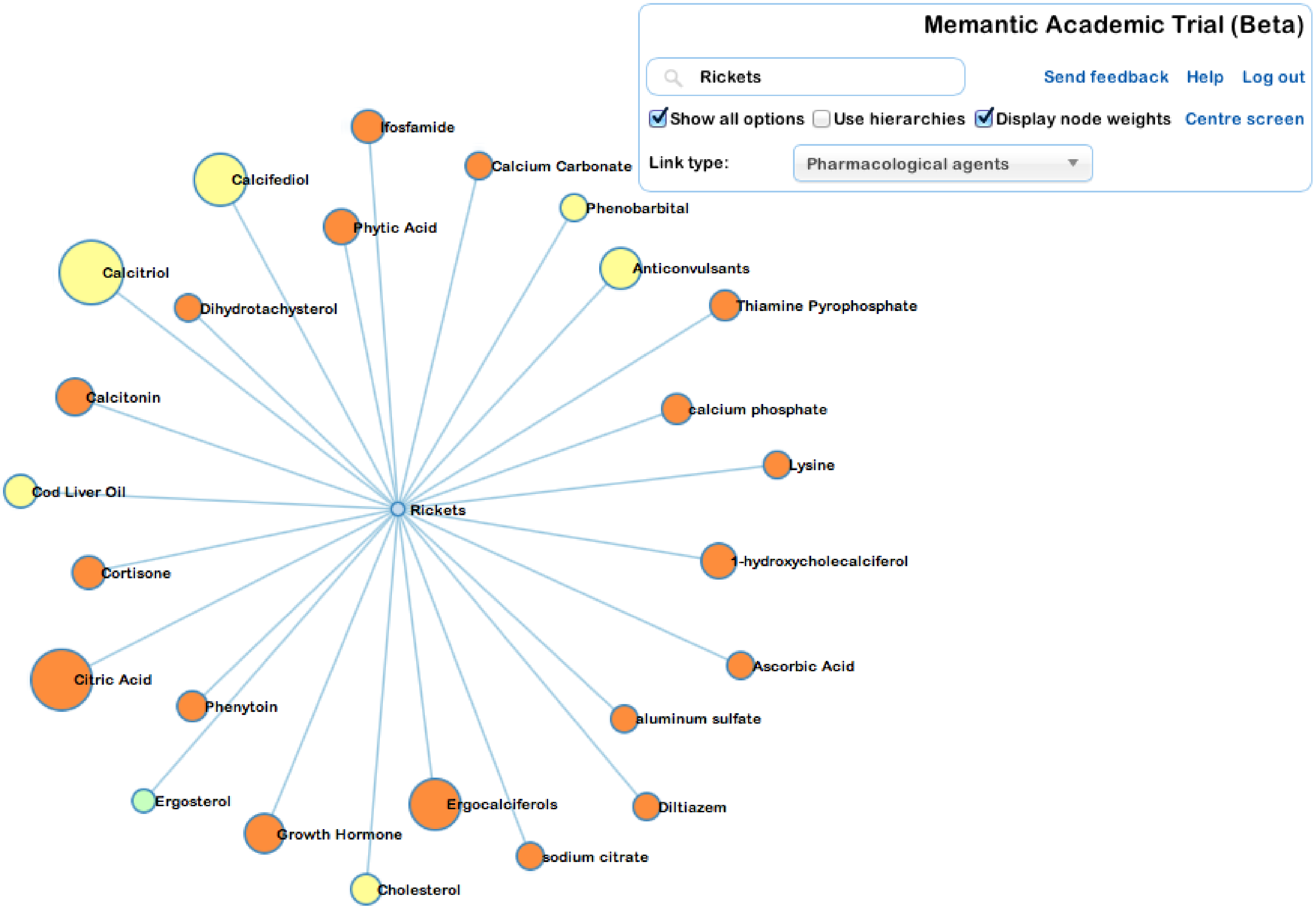}
\caption{Pharmacological agents associated with \emph{rickets}.}
\label{fig:filter_type}
\end{figure}

Memantic offers the ability to filter the visualised concepts by their type, such as ``diseases'', ``pharmacological agents'', ``therapeutic procedures'', and so on. For example, Figures \ref{fig:result_alz}, \ref{fig:result_toxo} and \ref{fig:result_alz_expand} display only diseases that are connected to the user's queries, which is the default setting. The link type can be selected in the drop down box in the floating toolbox. By way of example, Figure \ref{fig:filter_type} shows the set of pharmacological agents discovered by Memantic to be associated with rickets. Memantic uses the ``semantic type" field of MeSH dictionary entries to enable filtering concepts in the above manner; any semantic type present in MeSH can be used for filtering purposes.

\subsubsection{Leaf node colours.} Orange nodes signify concept associations that were discovered only in the scientific articles indexed from PubMed. Green nodes represent associations that were found only in medical encyclopaedias, and yellow nodes are used for associations that have been identified in both data sources.

\subsubsection{Node size.} 
\begin{figure}
\centering
\includegraphics[width=1\textwidth]{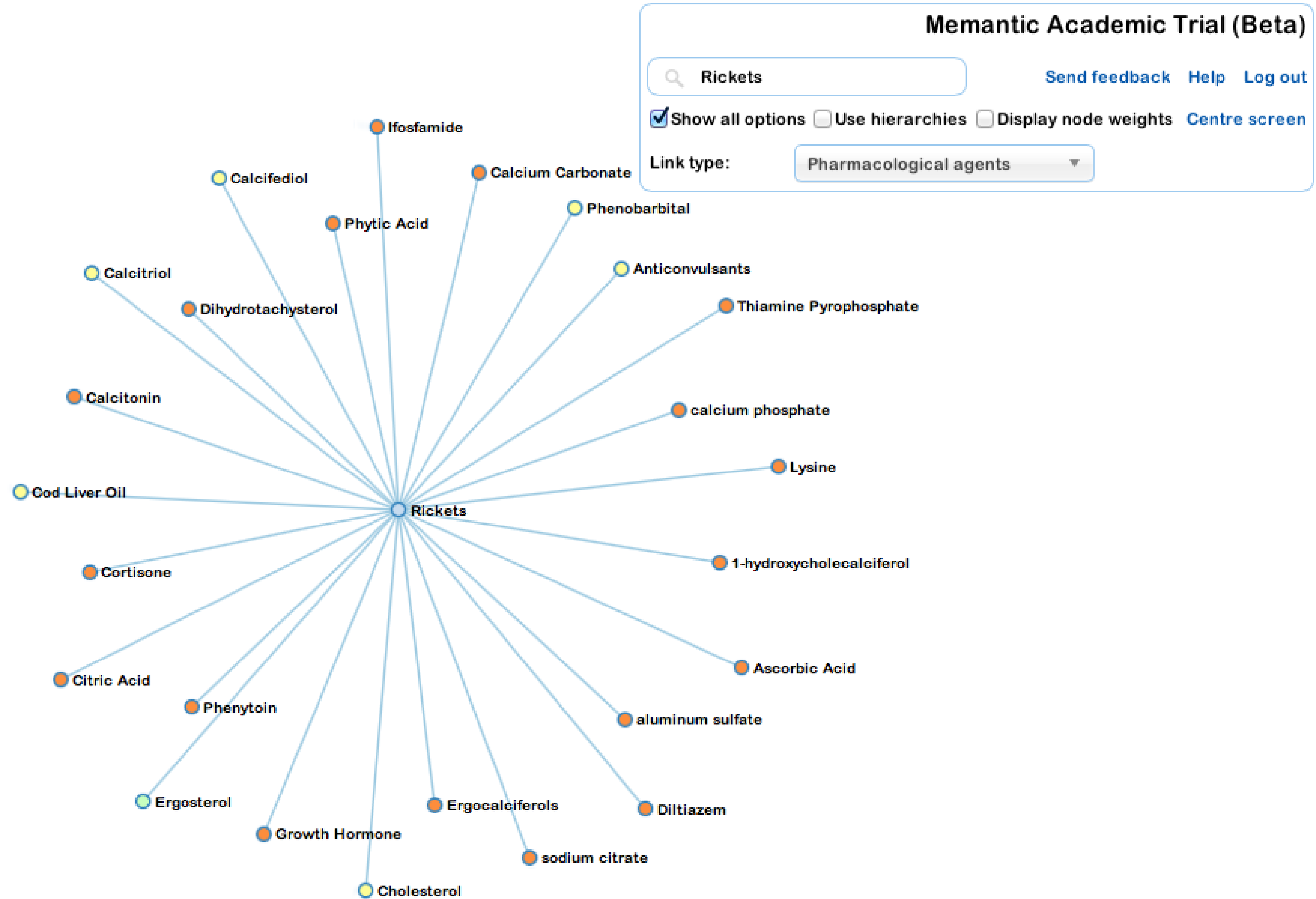}
\caption{Query results from Figure \ref{fig:filter_type} with node weighting turned off.}
\label{fig:noweights}
\end{figure}

The size of a leaf node (or its ``weight") is a function of the number of publications supporting the association of its concept with the query. The size of an intermediate node represents the sum of the sizes of all leaf nodes that descend from that node. This is a quick way to indicate the amount of available research for a particular association. It is possible to switch off the node weighting by toggling the ``node weights" checkbox (Figure \ref{fig:noweights}).

\subsubsection{Query box.} 
\begin{figure}
\centering
\includegraphics[width=0.65\textwidth]{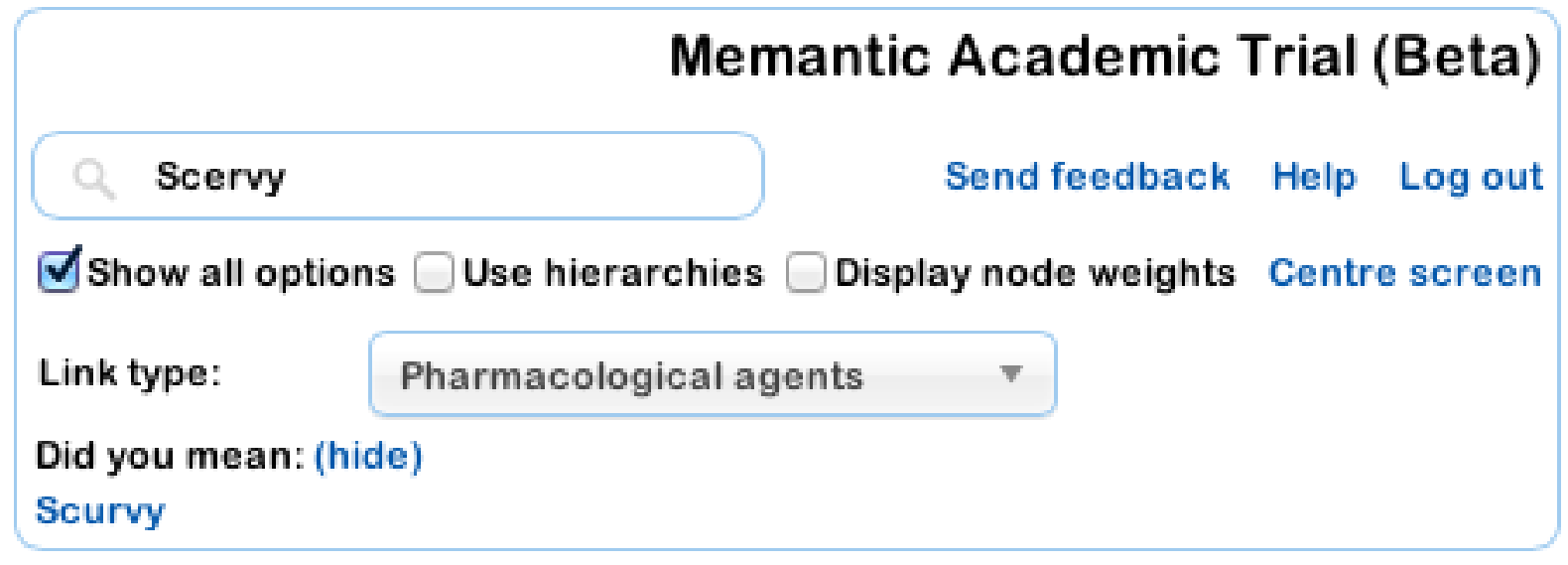}
\caption{Floating toolbox displaying suggested corrections for a new query with a spelling mistake.}
\label{fig:toolbox}
\end{figure}

Subsequent queries can be entered in the toolbox in the top right corner of the screen. The toolbox also has the facility to offer query suggestions and to submit user feedback (Figure \ref{fig:toolbox}).

\subsection{Demonstration videos}

In addition to the figures we present in this paper, we have created two demo videos. The first video, located at \url{http://youtu.be/201yfgNKzzc}\footnote{Also available at \url{http://vimeo.com/122447677} and \url{http://archive.org/details/MemanticAlzheimer}} follows a user issuing a query for Alzheimer disease, as shown in Figures \ref{fig:query} and \ref{fig:result_alz}, navigating to the desired disease connections through the medical concept hierarchy (Figure \ref{fig:result_alz_expand}), and exploring one connection that is only present in the PubMed database (Figure \ref{fig:result_alz_pubs}) and one that is also present in the Medscape medical encyclopaedia (Figure \ref{fig:result_alz_pubs_enc}).

The second video, uploaded to \url{http://youtu.be/fH7BMVBaLZc}\footnote{Also available at \url{http://vimeo.com/122451794} and \url{http://archive.org/details/MemanticRickets}}, illustrates the same user issuing a query for rickets, changing the link type to  ``pharmacological agents", switching to flat visualisation mode (Figure \ref{fig:filter_type}), and then toggling the ``node weights'' option. 

\section{Case studies} \label{sec:cases} 

\subsection{Broken heart syndrome, a.k.a. Takotsubo cardiomyopathy}

Takotsubo cardiomyopathy is a temporary condition where one's heart muscle becomes suddenly weakened. Also called acute stress cardiomyopathy, broken heart syndrome and apical ballooning syndrome, the condition causes the heart's left ventricle to change shape. According to the British Heart Foundation, the main symptoms are chest pains and breathlessness, similar to a heart attack. However, the key difference from the latter condition is the absence of any blockages in the coronary arteries, as confirmed by the angiogram imaging test. Although the cause of this condition has not been confirmed, it has been reported that approximately 75\% of people diagnosed with Takotsubo cardiomyopathy have experienced significant emotional or physical stress prior to the onset of the symptoms. There is some evidence that the excessive release of hormones (particularly adrenaline) during such periods of stress causes the weakening of the heart muscle \cite{bhf:takotsubo}.

\begin{figure}
\centering
\includegraphics[width=1\textwidth]{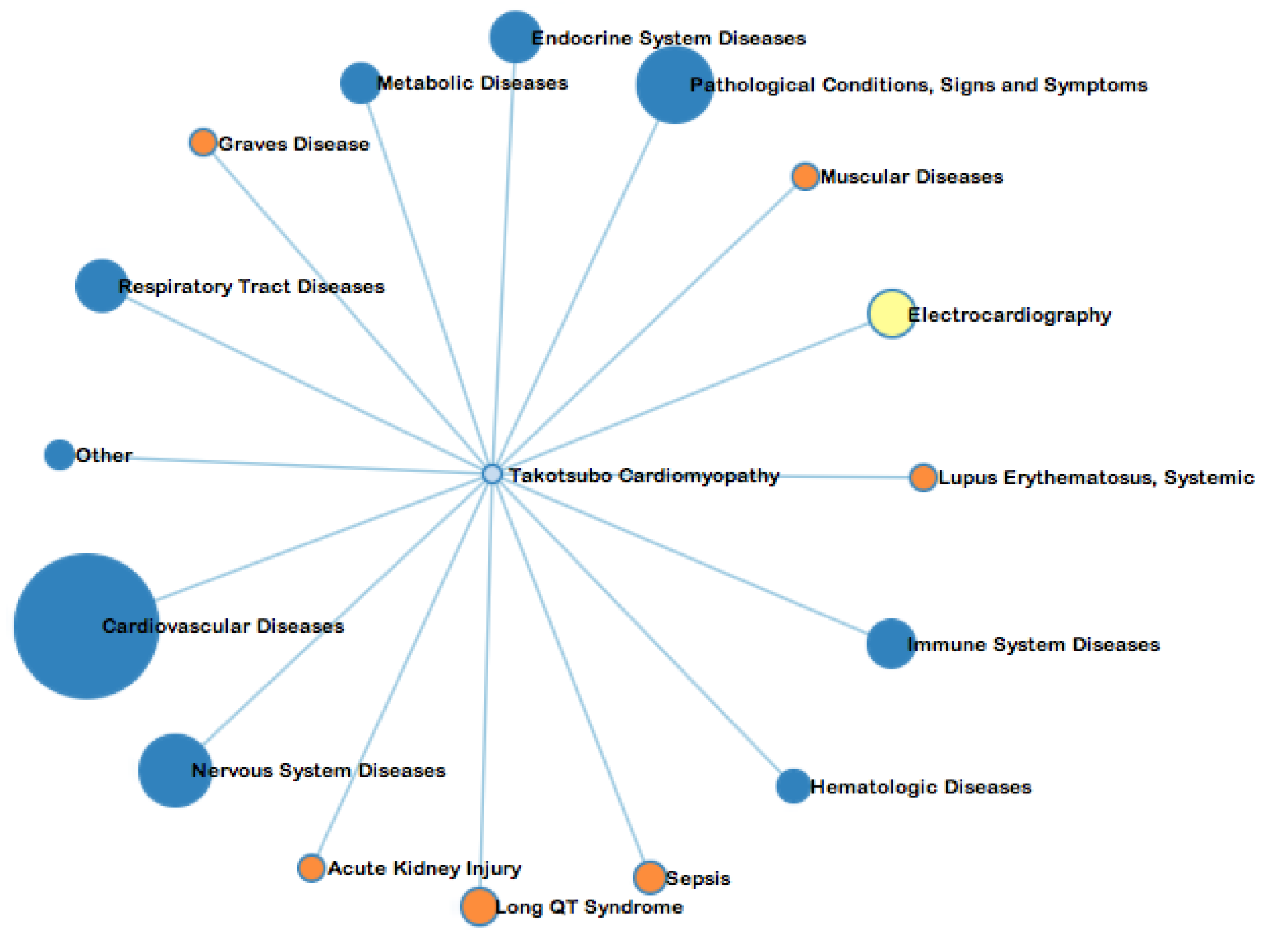}
\caption{Diseases related to \emph{Takotsubo cardiomyopathy}.}
\label{fig:tako}
\end{figure}

\begin{figure}
\centering
\includegraphics[width=1\textwidth]{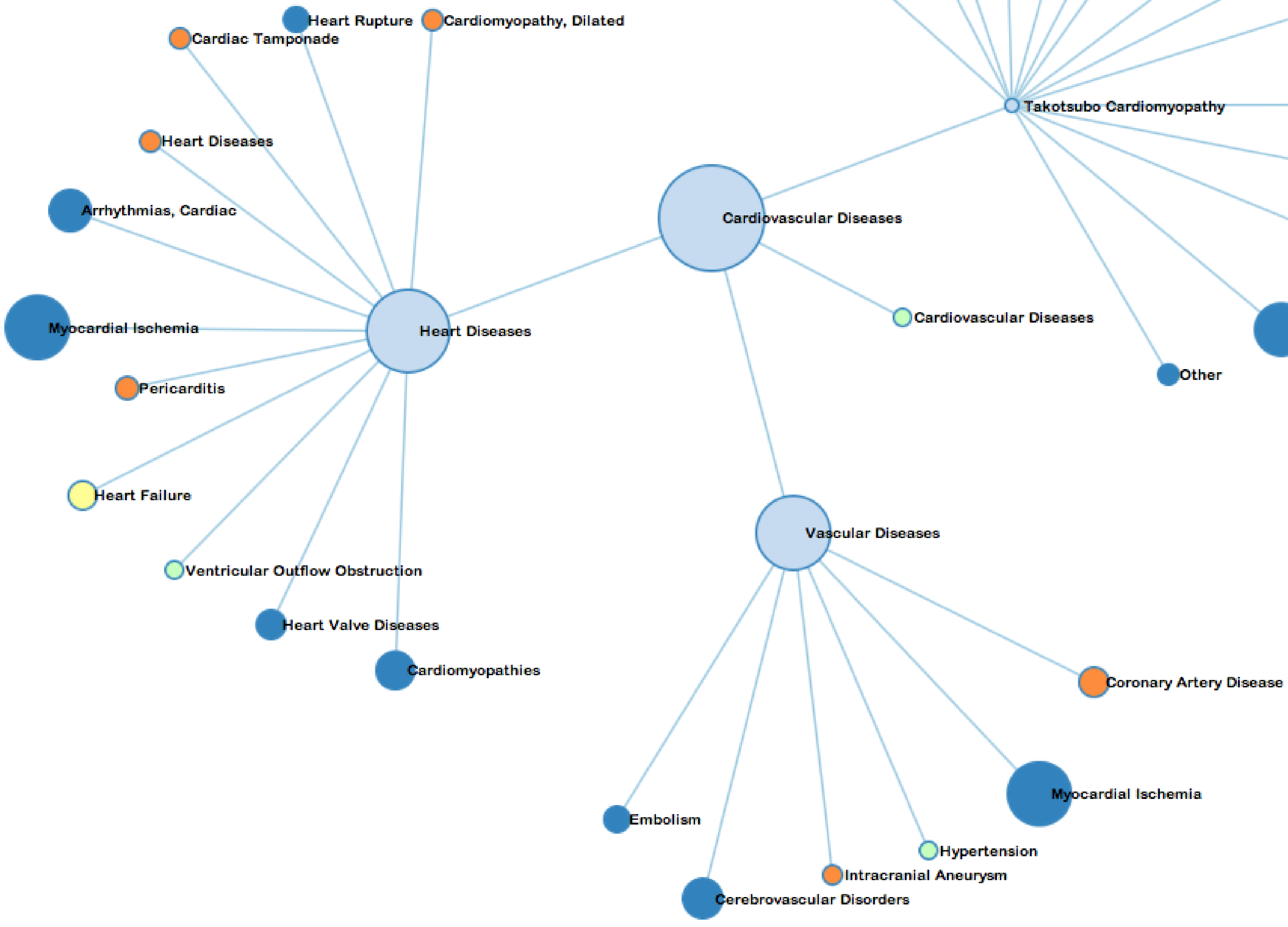}
\caption{Cardiovascular diseases related to \emph{Takotsubo cardiomyopathy}.}
\label{fig:tako_cardio}
\end{figure}

\begin{figure}
\centering
\includegraphics[width=1\textwidth]{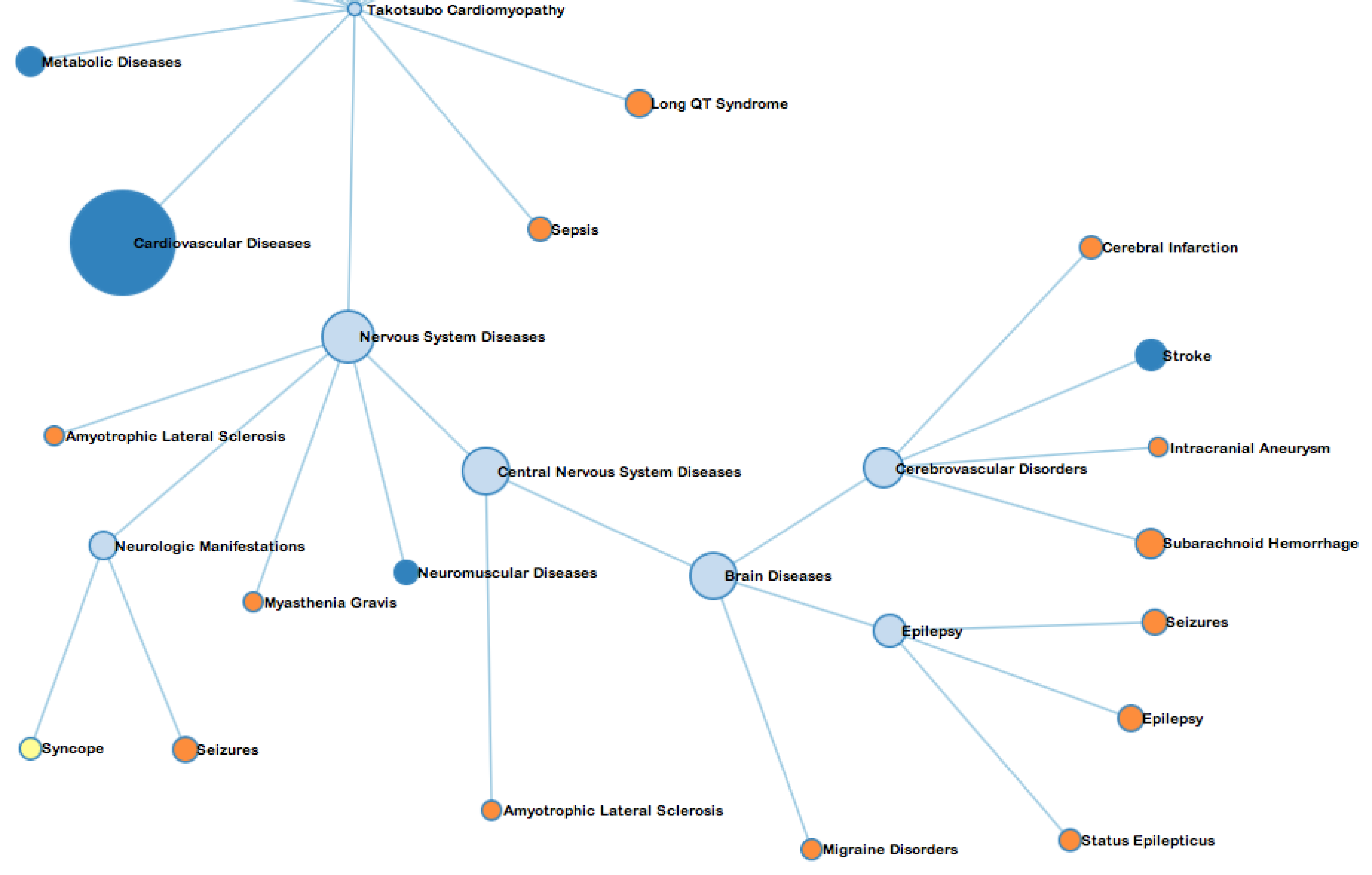}
\caption{Nervous system diseases related to \emph{Takotsubo cardiomyopathy}.}
\label{fig:tako_neuro}
\end{figure}

\begin{figure}
\centering
\includegraphics[width=1\textwidth]{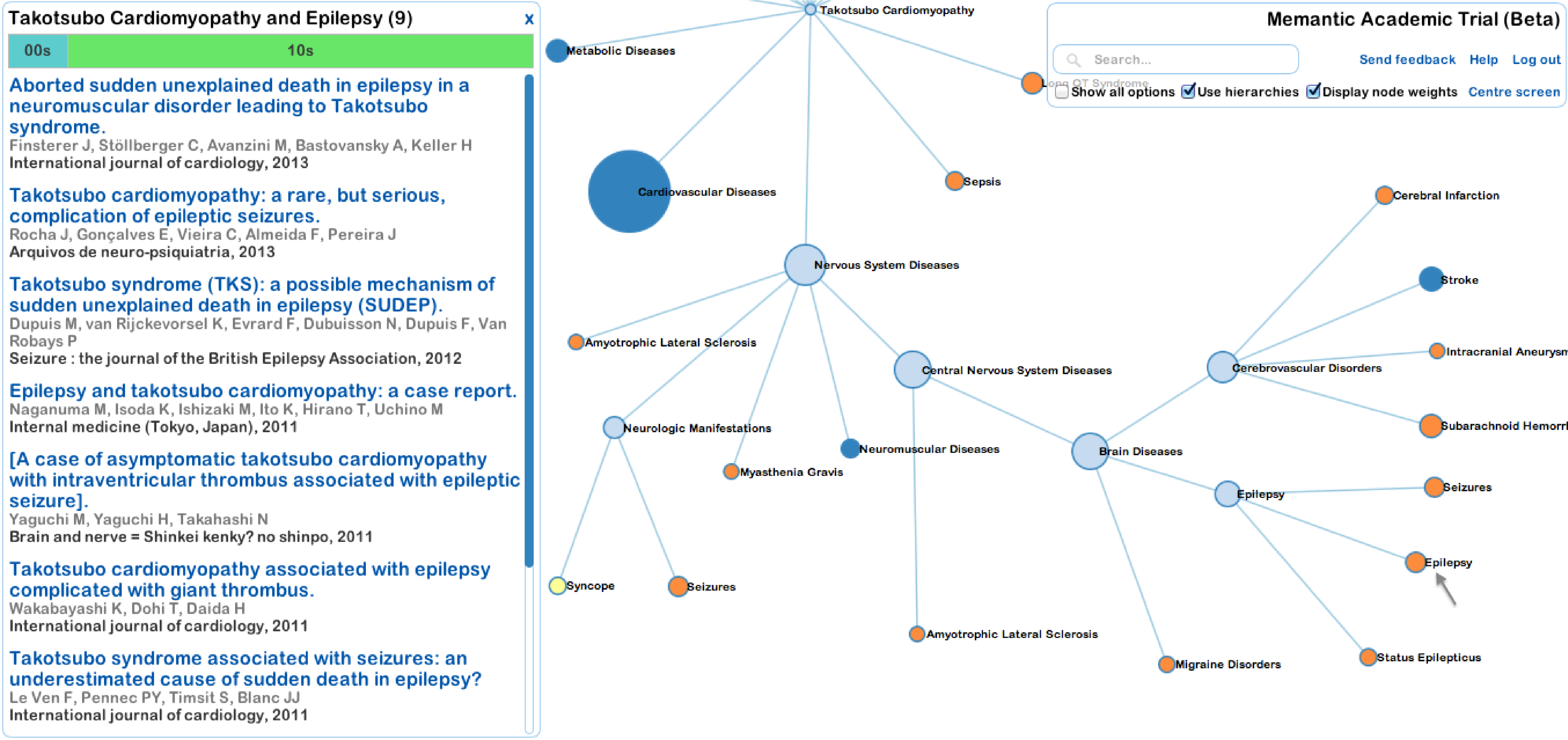}
\caption{Exploring the connection between \emph{Takotsubo cardiomyopathy} and \emph{epilepsy}.}
\label{fig:tako_epilepsy}
\end{figure}

At the time of writing, the pages about Takotsubo cardiomyopathy on the websites of the Mayo Clinic \cite{mayo:takotsubo}, Wikipedia \cite{wiki:takotsubo} and Medscape \cite{medscape:takotsubo} echoed the above description and cited psychological and physical stress as the main triggers for the syndrome. Memantic reveals a somewhat more detailed picture (Figure \ref{fig:tako}). It is immediately apparent that the largest group of associated diseases and syndromes are of cardiovascular nature, which is quite expected (Figure \ref{fig:tako_cardio}). However, another large cluster represents nervous system diseases, of which there is little or no mention in the above online resources (Figure \ref{fig:tako_neuro}). By way of example, exploring the connection between Takotsubo cardiomyopathy and epilepsy shows that most papers investigating the possibility of the latter being a trigger have been published in the current decade (Figure \ref{fig:tako_epilepsy}). Figure \ref{fig:tako_neuro} shows that most concept nodes in this subcategory are orange, which means that the associated conditions were not listed in the Medscape medical encyclopaedia when Memantic had last indexed it. After reviewing the articles in the nervous system disease category one begins to get the idea that brain disorders may also be involved in the hormonal imbalances that can lead to this type of cardiomyopathy.

\subsection{Early diagnosis of Alzheimer disease by retinal examination}

\begin{figure}
\centering
\includegraphics[width=1\textwidth]{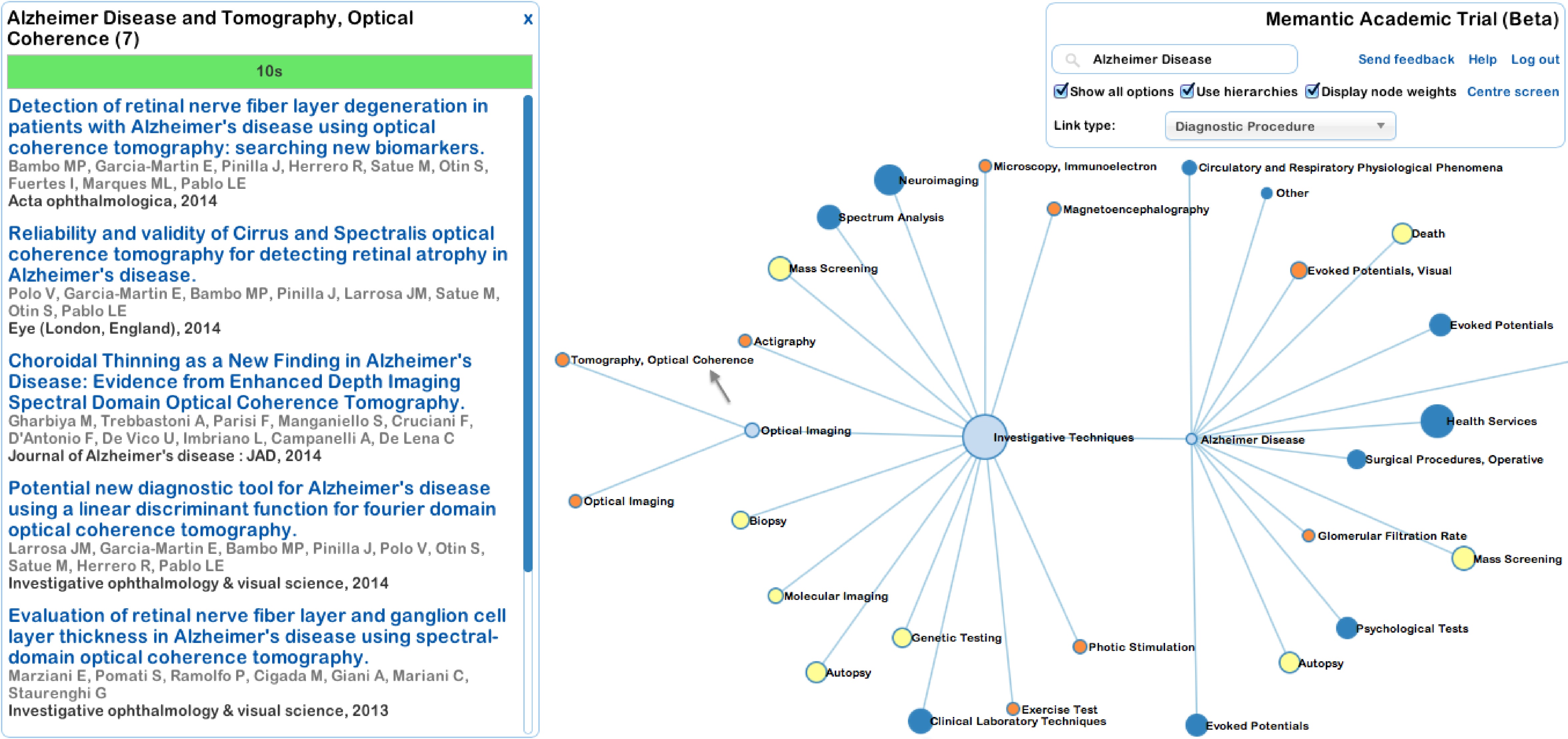}
\caption{The connection between \emph{Alzheimer disease} and \emph{optical coherence tomography}.}
\label{fig:alz_oct}
\end{figure}

Alzheimer disease is a neurodegenerative disorder characterised by a decline in cognitive function. It mainly affects the elderly and while there are currently no known treatments that are effective, accurate early diagnosis may prove critical for possible future intervention approaches. Figure \ref{fig:alz_oct} shows Memantic visualising diagnostic procedures related to Alzheimer disease. In particular, it shows the recent research on the use of Optical Coherence Tomography (a retinal imaging technique) for identifying the onset of the condition. Note that the orange colour of the associated node indicates that this information was not present in the Medscape medical encyclopaedia when Memantic had last indexed it.

\subsection{Exploring the association between measles and multiple sclerosis}

\begin{figure}
\centering
\includegraphics[width=1\textwidth]{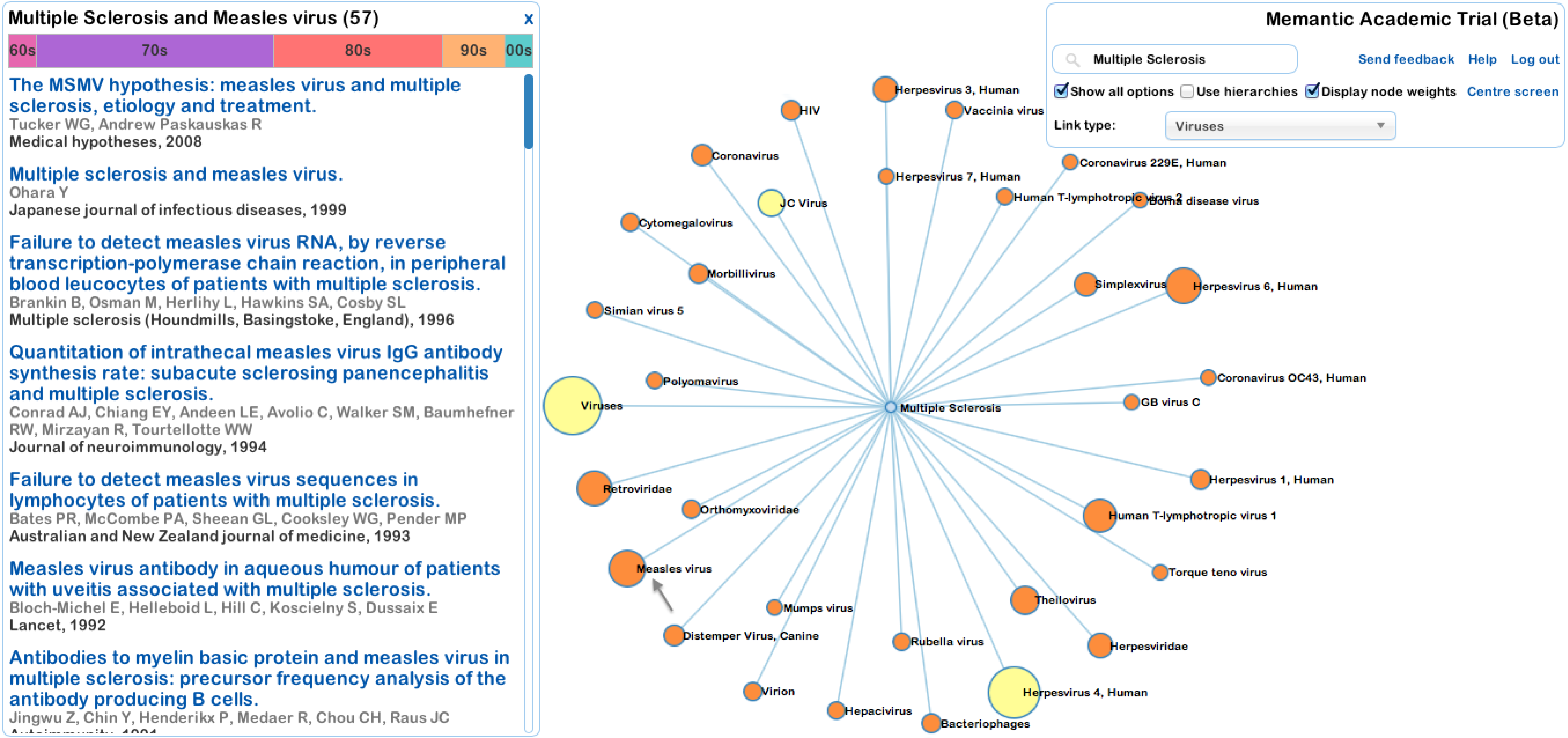}
\caption{The connection between \emph{multiple sclerosis} and the \emph{measles virus}.}
\label{fig:ms_measles}
\end{figure}

	Multiple sclerosis (MS) is a disease in which the immune system attacks the protective sheath (myelin) that covers nerve fibres. Damage to myelin disrupts the electrical signals that are passed along the nerves between the brain and the rest of the body. This disruption can result in a variety of neurological symptoms, such as the loss of motor skills and vision. Little is currently known about what triggers the immune system to attack myelin, but at one point it was thought that exposure to the measles virus could play a role. Figure \ref{fig:ms_measles} shows Memantic's visualisation of viruses related to MS and the list of publications concerning the relationship between MS and the measles virus. The stacked bar chart that breaks down publication numbers by decade indicates that this hypothesis was investigated most intensively in the 1970s and 80s and is no longer actively pursued.

\subsection{The link between asthma and vitamin D deficiency}

\begin{figure}
\centering
\includegraphics[width=1\textwidth]{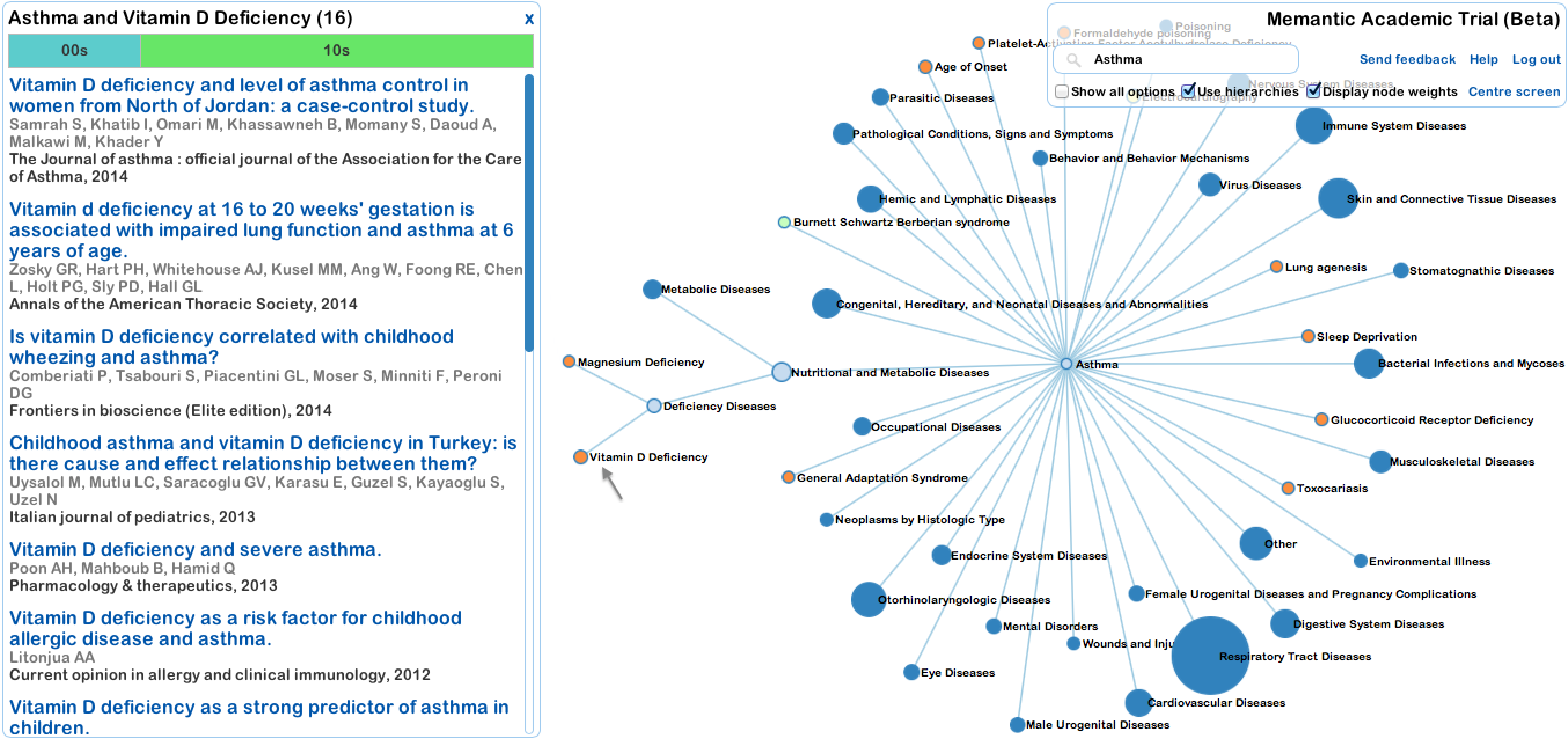}
\caption{The connection between \emph{asthma} and \emph{vitamin D deficiency}.}
\label{fig:asthma_vitd}
\end{figure}

Asthma is a disease of the respiratory system and manifests itself through recurrent episodes of airway constriction. It is considered a complex and multifactorial disease, since a number of different genes are thought to play a role in the disorder and a variety of environmental factors can act as triggers. Figure \ref{fig:asthma_vitd} shows Memantic's visualisation of diseases related to asthma and the list of publications concerning the relationship of the disease to vitamin D deficiency. The decade bar chart indicates that this research is relatively recent and the orange colour of the associated node implies that this relationship is not mentioned in the Medscape medical encyclopaedia's article on asthma. 

\subsection{Mental health conditions associated with the use of cannabis}

An increasingly active debate about legalising cannabis has been taking place in recent decades in much of the western world. The potential mental health risks associated with the psychoactive substances present in the drug are a major part of this debate. Figure \ref{fig:asthma_vitd} shows Memantic's visualisation of the mental and behavioural dysfunctions related to cannabis and the list of publications concerning the relationship of the drug to psychosis. It is evident that most of the research concerning this relationship has been carried out in the last two decades. Schizophrenia, depression and bipolar disorder also feature prominently amongst other related mental health conditions.

\begin{figure}
\centering
\includegraphics[width=1\textwidth]{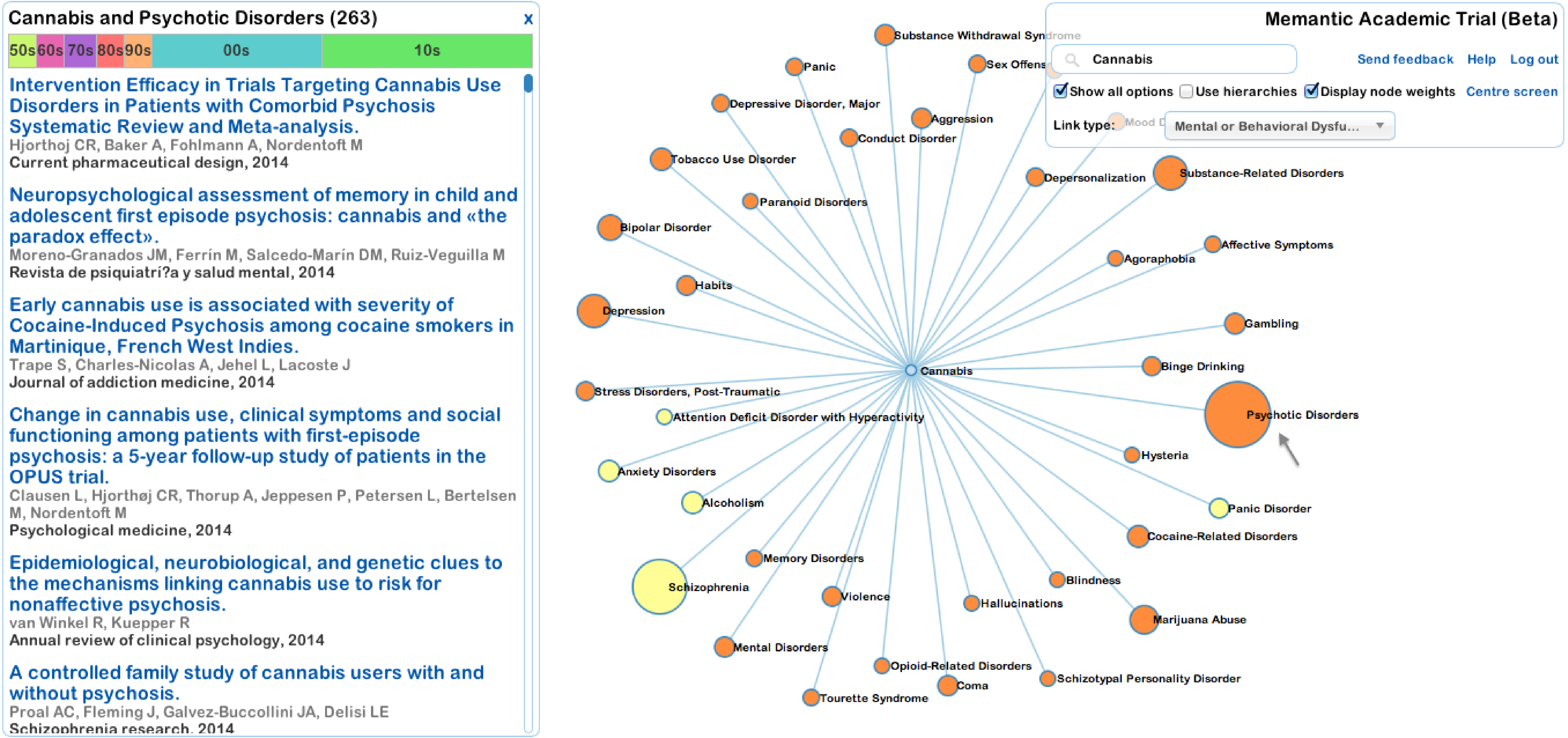}
\caption{The connection between \emph{cannabis} and \emph{psychosis}.}
\label{fig:cannabis}
\end{figure}

\section{Future work}

A significant drawback of our system is that queries can only contain one medical concept at a time. In the future, we would like to extend our approach to more complex queries, so that it would be possible to explore which concepts are related to a collection of medical terms, or to an arbitrary string of text that is not present in our medical dictionary. We would also like to offer users the ability to manually correct concept relationships that were extracted erroneously.

\section*{About the author}

Alexei Yavlinsky is a software engineer with a PhD in Computing from Imperial College London. His interests lie in the fields of machine learning, statistics and human biology. Prior to founding Behold Research, he worked for a number of London IT startups, specialising in applying data mining techniques.

\bibliography{typeinst}{}

\begin{thebibliography}{10}

\bibitem{grivell:2002}
L.~Grivell.
\newblock Mining the bibliome: searching for a needle in a haystack?
\newblock {\em EMBO Reports}, 3(3):200--203, March 2002.

\bibitem{lowe:1994}
H.~J. Lowe and G.~O. Barnett.
\newblock Understanding and using the medical subject headings ({MeSH})
  vocabulary to perform literature searches.
\newblock {\em JAMA}, 271(14):1103--1108, 1994.

\bibitem{callon:1983}
M.~Callon, J.~P. Courtial, W.~A. Turner, and S.~Bauin.
\newblock From translations to problematic networks: An introduction to co-word
  analysis.
\newblock {\em Social Science Information}, 22(2):191--235, 1983.

\bibitem{jacunski:2013}
A.~Jacunski and N.~P. Tatonetti.
\newblock Connecting the dots: applications of network medicine in pharmacology
  and disease.
\newblock {\em Clinical Pharmacology \& Therapeutics}, 94(6):659--669, 2013.

\bibitem{goh:2007}
K.~Goh, M.~E. Cusick, D.~Valle, B.~Childs, M.~Vidal, and A.~Barab{\'a}si.
\newblock The human disease network.
\newblock {\em Proceedings of the National Academy of Sciences},
  104(21):8685--8690, 2007.

\bibitem{weiss:1974}
S.~M. Weiss.
\newblock {\em A system for model-based computer-aided diagnosis and therapy}.
\newblock PhD thesis, Rutgers University, 1974.

\bibitem{weiss:1978}
S.~M. Weiss, C.~A. Kulikowski, and A.~Safir.
\newblock Glaucoma consultation by computer.
\newblock {\em Computers in Biology and Medicine}, 8(1):25--40, 1978.

\bibitem{shortliffe:1975}
E.~H. Shortliffe and B.~G. Buchanan.
\newblock A model of inexact reasoning in medicine.
\newblock {\em Mathematical biosciences}, 23(3):351--379, 1975.

\bibitem{shortliffe:1976}
E.~H. Shortliffe.
\newblock {MYCIN}: Computer-based medical consultations, 1976.

\bibitem{shortliffe:1981}
E.~H. Shortliffe, A.~C. Scott, M.~B. Bischoff, A.~B. Campbell, W.~van Melle,
  and C.~D. Jacobs.
\newblock {ONCOCIN}: An expert system for oncology protocol management.
\newblock In {\em Proceedings of the 7th International Joint Conference on
  Artificial Intelligence}, pages 876--881, 1981.

\bibitem{pople:1975}
H.~E. Pople, J.~D. Myers, and R.~A. Miller.
\newblock {DIALOG}: a model of diagnostic logic for internal medicine.
\newblock In {\em Proceedings of the 4th International Joint Conference on
  Artificial Intelligence}, pages 848--855. Morgan Kaufmann Publishers Inc.,
  1975.

\bibitem{miller:1982}
R.~A. Miller, H.~E. Pople, and J.~D. Myers.
\newblock {INTERNIST-I}, an experimental computer-based diagnostic consultant
  for general internal medicine.
\newblock {\em The New England Journal of Medicine}, 307(8):468, 1982.

\bibitem{barnett:1987}
G.~O. Barnett, J.~J. Cimino, J.~A. Hupp, and E.~P. Hoffer.
\newblock {DXplain}: an evolving diagnostic decision-support system.
\newblock {\em JAMA}, 258(1):67--74, 1987.

\bibitem{dxplain:online}
{DXplain}: Using decision support to help explain clinical manifestations of
  disease.
\newblock \url{http://www.mghlcs.org/projects/dxplain}.

\bibitem{mapofmed:online}
{Map of Medicine}: Healthcare management solutions.
\newblock \url{http://mapofmedicine.com}.

\bibitem{watpaths:online}
{WatsonPaths}: A new cognitive computing project that enables more natural
  interaction between physicians, data and electronic medical records.
\newblock
  \url{http://www.research.ibm.com/cognitive-computing/watson/watsonpaths.shtm%
l}.

\bibitem{watpaths:schools}
{IBM Research} unveils two new {Watson} related projects from cleveland clinic
  collaboration.
\newblock \url{http://www-03.ibm.com/press/us/en/pressrelease/42203.wss}.

\bibitem{ghealthcards:online}
{Google Operating System Blog}: Google health cards.
\newblock
  \url{http://googlesystem.blogspot.co.uk/2013/11/google-health-cards.html}.

\bibitem{deerwester:1990}
S.~C. Deerwester, S.~T. Dumais, T.~K. Landauer, G.~W. Furnas, and R.~A.
  Harshman.
\newblock Indexing by latent semantic analysis.
\newblock {\em Journal of the American Society for Information Science},
  41(6):391--407, 1990.

\bibitem{koshman:2006}
S.~Koshman, A.~Spink, and B.~J. Jansen.
\newblock Web searching on the {Vivisimo} search engine.
\newblock {\em Journal of the American Society for Information Science and
  Technology}, 57(14):1875--1887, 2006.

\bibitem{pubmed:online}
{PubMed}.
\newblock \url{http://www.ncbi.nlm.nih.gov/pubmed}.

\bibitem{medscape:online}
{Medscape}.
\newblock \url{http://www.medscape.com}.

\bibitem{d3:online}
{D3.js} \--- data-driven documents.
\newblock \url{http://d3js.org}.

\bibitem{bhf:takotsubo}
{British Heart Foundation \--- Takotsubo cardiomyopathy}.
\newblock
  \url{https://www.bhf.org.uk/heart-health/conditions/cardiomyopathy/takotsubo%
-cardiomyopathy}.

\bibitem{mayo:takotsubo}
{The Mayo Clinic \--- Takotsubo cardiomyopathy}.
\newblock \par
  \url{http://www.mayoclinic.org/diseases-conditions/broken-heart-syndrome/bas%
ics/definition/con-20034635}.

\bibitem{wiki:takotsubo}
{Wikipedia \--- Takotsubo cardiomyopathy}.
\newblock \par \url{http://en.wikipedia.org/wiki/Takotsubo_cardiomyopathy}.

\bibitem{medscape:takotsubo}
{Medscape \--- Takotsubo cardiomyopathy}.
\newblock \par \url{http://emedicine.medscape.com/article/1513631-overview}.

\end{thebibliography}
\bibliographystyle{unsrt}

\end{document}